\documentclass{iopart} % final
\usepackage{iopams}

\usepackage{graphicx}
\graphicspath{{pic/}}

\usepackage{color}

\begin{document}

\title{Optical Detector Topology for Third-Generation Gravitational Wave Observatories}


\author{A~Freise$^1$, S~Hild$^3$, K~Somiya$^2$, K~A~Strain$^3$, A~Vicer\'e$^4$, M~Barsuglia$^5$ and S~Chelkowski$^1$}

\address{$^1$   School of Physics and Astronomy, University of Birmingham, 
              Edgbaston, Birmingham B15 2TT, UK}
\address{$^2$ Theoretical Astrophysics, California Institute of Technology, 
         Pasadena, California, 91125, US}
\address{$^3$  Institute for Gravitational Research, University of Glasgow,
             Glasgow G12 8QQ, UK}
\address{$^4$  Universit\'a degli Studi di Urbino "Carlo Bo" and INFN -
         Sezione di Firenze, Italy}
\address{$^5$ AstroParticule et Cosmologie, CNRS UMR7164, Paris, France}

%\maketitle

\begin{abstract}
The third generation of gravitational wave observatories, aiming to  provide
100 times better sensitivity than currently operating interferometers,
is expected to establish the evolving field of gravitational wave
astronomy. A key element for achieving the ambitious sensitivity
goal is the exploration of new interferometer geometries, topologies
and configurations.  In this article we review the current status of
the ongoing design work
for third-generation gravitational wave observatories. The main focus
is set on the evaluation of the detector geometry and detector topology. In addition
we discuss some promising detector configurations and potential
noise reduction schemes.

%\keywords{gravitational wave detector \and laser interferometer \and topology \and optical design}
\pacs{04.80.Nn, 07.60.Ly, 95.75.Kk, 95.55.Ym}
\end{abstract}

\section{Introduction}
\label{intro}
In the early stages of a design process towards third-generation gravitational wave detectors we can
indulge in the creative activity of inventing completely new instruments in various
forms or shapes.
However, third-generation detectors would be very large (several kilometers long) instruments and
we can expect that in case of a successful realisation
the final form and location of the detectors will have been strongly influenced by less scientific constraints
such as costs and  politics. The most exciting period for many instrumental designers lies in the period
when out of the many possibilities we have to identify and optimise the best instrument
conceivable within an optimistic but realistic outlook on the available resources and technologies.

\section{The footprint of future interferometric detectors}
\label{sec:adf}

This section briefly reviews the reasoning behind the shape of current gravitational wave detectors and then
discusses alternative geometries which can be of interest for third-generation detectors. We
will use the terminology introduced in the review of a triangular configuration~\cite{Freise09} and
discriminate between the
\emph{geometry}, \emph{topology} and \emph{configuration} of a detector as follows:
\begin{itemize}
\item The \emph{geometry} describes the position information of one or several interferometers,
defined by the number of interferometers, their location and relative orientation.
\item The \emph{topology} describes the optical system formed by its core elements.
The most common examples are the Michelson, Sagnac and Mach--Zehnder topologies.
\item Finally the \emph{configuration} describes the detail of the optical layout and the set of parameters that
can be changed for a given topology, ranging from the specifications of the optical core
elements to the control systems, including the operation point of the main interferometer.\footnote{Note that the
addition of optical components to a given topology is often referred to as a change in configuration.}
\end{itemize}
%%% TODO: the above could do with a bit of a rewrite

\subsection{The L-shape}
Current gravitational wave detectors represent the most precise instruments for measuring length changes.
They are laser interferometers with km-long arms and are operated differently from many precision
instruments built for measuring an absolute length. Viewed from above
they resemble an L-shape with equal arm length.
This geometric form follows directly from the nature of gravitational
waves: gravitational waves are transverse, quadrupole waves, thus a length change measured along any axis
occurs with opposite sign along the axis orthogonal to the previous and the direction of propagation.
This key feature allows to make a differential measurement between two orthogonal interferometer
arms, yielding twice the amplitude of a single arm. More importantly a differential measurement allows us to
potentially discriminate between gravitational wave signals and those types of noise common to both arms, such as,
for example, laser amplitude noise. To achieve this the interferometer arms generally have to have approximately
the same length. The most simple L-shaped interferometer allowing to do this type
of measurement is the symmetric Michelson interferometer, on whose topology all current interferometric detectors
are based.

The long arm length of the detectors represents the simplest way to increase the signal-to-noise ratio
in the detector because the `tidal' effect of the gravitational wave increases with the base length over which the
measurement is taken, while the fundamental noises are connected to the interaction
of light with the optical components or the photo detection and thus do not scale with the length of the interferometer
arms.
%%% adf 31.01.2010 changed the text below because of reviewers 2 comments
We can summarise, provided specifications of the vacuum system housing the interferometer and
the performance of mirror position control systems are good enough, an increase in arm length will increase the
sensitivity of the detector proportionally.

Using the framework developed in~\cite{Jaranowski98} we can compute the sensitivity of a laser interferometer
with two arms to gravitational waves, taking into account the geometry of the detector, the location of the source and the
changes of both over time. The equations show directly that the arms of the detector do not
have to be perpendicular\footnote{The GEO detector for example features an opening angle of approximately $94^\circ$
in order to make the best use of the available site.}, the right angle, however, provides the maximum
response of an ideal detector to gravitational waves, which more generally can be written as
\begin{equation}
h(t)=F_{+}(t)h_{+}(t)+F_{\times}(t)h_{\times}(t)=\sin\zeta\,f(t,\psi, \dots)
\end{equation}
with $\zeta$ the opening angle of the interferometer arms, $F_{+}$ and $F_{\times}$ the beam pattern functions
and $f(t,\psi, \dots)$ a functions of the remaining parameters describing the geometry (the location of the detector and
of the source in space and time and the wave polarisation angle).

In summary we can say that for a gravitational wave of given direction and polarisation, a properly aligned  symmetric L-shape
is an ideal optical layout for an interferometric
detector; the arms should be as long as possible and the sensitivity is maximised for an opening angle of $90^\circ$.
It should be noted that this does not put severe constraints on the type of interferometer topology used. In fact,
most common interferometer types can be used in a form that features two large symmetric arms in an L-shape
while potential other interferometer arms or sections are shortened such that they can be considered as part of one
corner of the detector.
%%% TODO: maybe add figure? A few example layouts are shown in Figure~\ref{fig:L-layouts} to illustrate this.

\subsection{The triangle}
\begin{figure}[tbh]
\begin{center}
\includegraphics[width=0.7\textwidth,keepaspectratio]{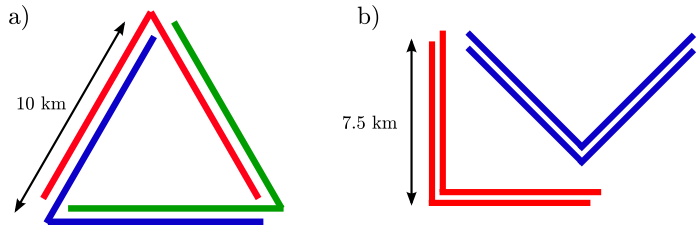}
\caption{{\bf a)} Triangle geometry: three L-shaped detectors with 10\, km arm length
are positioned in a equilateral triangle.
{\bf b)} Four L-shaped detectors at $0^\circ$ and $45^\circ$. The integrated length of all
interferometer arms in both configurations is 60\,km and two interferometer arms can share
the same structure. Note that for avoiding noise correlations between two detectors the
neighbouring interferometer arms would probably be housed in a separate vacuum tubes.}
\label{fig:triangle}
\end{center}
\end{figure}

At any given moment an L-shaped detector can only detect one linear combination of polarisations of a
gravitational wave. However, for
estimation of source parameters from the measured signal, the full polarisation information can be essential
(see next section). %%% TODO check this cross reference!!
Thus it is of considerable interest to design a detector able to detect both polarisations (and thus the full content)
of a gravitational wave at all times. This can be achieved by combining two co-located
L-shaped detectors which are positioned at $45^\circ$ to each other. % (as shown in Figure~\ref{fig:triangle}).
Already more than 20 years ago it was recognised that a triangular geometry would provide the same
sensitivity to both polarisations as detectors at $45^\circ$ while requiring less enclosed space and fewer
end stations~\cite{MPQ-talk}. In particular, the sensitivity of the two geometries shown
in Figure~\ref{fig:triangle} differs only by $6\%$~\cite{Freise09}. The difference in the sensitivity to
different polarisations between a single L-shape and a triangular geometry can be best illustrated
with a plot of the so-called antenna pattern as shown in Figure~\ref{fig:triangleAP}.

Using co-located detectors yields another advantage. Both layouts shown in Figure~\ref{fig:triangle} represent
detectors with redundancy. Redundancy here can be understood in relation to the continuous operation of the
detector as an observatory, or as a feature of the data streams generated by the full system. Redundancy in operation
is achieved by having multiple detectors which generate an equal or similar response to gravitational waves.
This is desirable in observatories which are expected to produce a quasi-continuous stream of astrophysical meaningful
data over an substantial amount of time. Typically laser interferometers cannot produce science data during
upgrades and maintenance work. Thus only alternate upgrading and data taking of redundant detectors can
avoid long down-times, for example during detector upgrades.

Such redundancy is obviously provided in the case of the 4 L-shaped  detectors, where two detectors are
always identical but can be operated independently. However, one can easily show that the triangular geometry
provides exactly the same redundancy~\cite{Freise09}. For example, for three equal L-shaped interferometers
oriented at $0^\circ$, $120^\circ$ and $240^\circ$, one obtains:
\begin{equation}
-h_{0^\circ}= h_{240^\circ}+h_{120^\circ}\,,
\end{equation}
where the sign of the operation is defined by which ports of the interferometers
are used to inject the laser light. Thus the two interferometers at $120^\circ$ and $240^\circ$
create exactly the same response as the one at $0^\circ$. This allows to construct
so-called null-streams (or null-data streams)~\cite{nullstream}. Null-streams are a powerful
data analysis method that allows to identify noise which is uncorrelated between the
detectors. Even though this does not increase the sensitivity of a detector,
it can add significantly to the robustness of the data processing pipelines and thus lead, for example, to shorter
delays between an event and the generation of a trigger for follow-up searches with optical telescopes.
The triangular geometry represents the minimal setup in one plane that can resolve both polarisations
and provides redundancy for the generation of null-streams.

The idea of using a triangular geometry is considered with strong interest within the context of
the design study for a third-generation detector \emph{Einstein gravitational wave Telescope}~\cite{ET}.
Therefore we will in the following use the term \emph{ET-class} to refer to three L-shaped detectors in a
triangular geometry.

\begin{figure}[tbh]
\begin{center}
\includegraphics[width=0.25\textwidth]{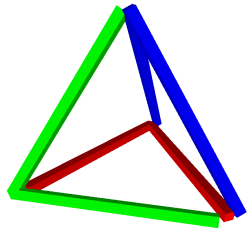}
\caption{The minimal three-dimensional geometry from L-shaped detectors
is an equilateral pyramid in which each edge houses one interferometer arm. A redundant set
which extends the features of the triangular shape requires six L-shaped detectors
such that each edge of the pyramid houses two detector arms.}
\label{fig:pyramid}
\end{center}
\end{figure}

\subsection{The pyramid}
When all co-located interferometers of one detector are confined to one plane they provide maximum sensitivity for
gravitational wave travelling perpendicular to this plane but are not very efficient for gravitational waves
travelling parallel to the plane. In particular, if the polarisation of the wave is not aligned with the
plane of the detectors the sensitivity can drop to zero. This, for example, is illustrated in the center
antenna pattern in Figure~\ref{fig:triangleAP}: in the central region (with detector normal orthogonal to
the z-axis) the sensitivity oscillates between zero and half of the maximum sensitivity as a function of
the angle of the detector normal to the polarisation axis.

In order to maximise the detector response regardless of the source location the collocated
detectors must form a three-dimensional structure. Ideally an redundant set of six detectors should
be used to preserve the features discusses for the triangular geometry. Such a detector would exhibit
an almost spherical antenna pattern as a whole and can provide unique source location information by comparing
the signals of the co-located detectors. The minimal approach to a full three-dimensional structure
requires three L-shaped detectors. Such a geometry is shown in Figure~\ref{fig:pyramid} and the
corresponding antenna pattern is depicted in the right plot of Figure~\ref{fig:triangleAP}.

While the pyramid can be considered the ideal geometry maximising the detected signal independent of the
source location or polarisation, this geometry has not been studied in detail. The reasons for this
are probably practical problems related to the non-horizontal light beams as well as the non-vertical
mirrors. The former at least would increase construction costs of the detector disproportionally
and the latter would even require a completely different suspension concept than employed by
current detectors.

%How shall these future detectors look like? The answer is: all interferometer ideas fit
%into various forms, we prefer the L-shape for GW detection or combinations of L-shapes.
%This summarizes the arguments from the `triple Michelson' paper and does
%same basic considerations about interferometer size etc, linked to the 'brute force' section below
\section{The ideal microphone}
\label{sec:andrea}

% \tcm{
% (maybe combined with or in relation to the previous section if necessary)
% The idea here is think how many interferometers per site we want and in what
% orientation. Thus, how can be build the perfect GW microphone. This
% could be considered for several example cases:}
% \begin{itemize}
% \item \tcm{many ETs distributed around the globe}
% \item \tcm{just one ET detector but several 'old' advanced detectors}
% \item \tcm{just one ET detector}
% \end{itemize}
% \tcm{I think a lot of this work still needs to be done but we could at least
% include simple examples and do a discussion about the problem
% where we don't know the answer yet. Maybe describe the methods
% and figures of merit etc.}

The third-generation gravitational wave detectors will be charged
with the mission of opening the field of precision physics in general
relativity~\cite{will:2006}, and provide data to the astrophysics community
allowing to complement information obtained by electro-magnetic and neutrino
detectors~\cite{gwic:2009}.
The details of the mission may depend on the results obtained with the second-detector generation, the so-called \emph{advanced detectors}, currently under
construction~\cite{vicere:2005}.
In fact, we can expect that advanced detectors will have not only carried out
the first detection, but collected a sufficient number of events to allow
identifying the most promising classes of sources, thus possibly suggesting
directions for the optimization of the design.

\subsection{The sources of interest}

For the third-generation ground based detectors to really allow precision
measurements on all the sources of interest, several requirements
with impact on the observatory geometry and topology need to be considered.
It seems appropriate then to recapitulate the characteristics of the different
classes of sources (see \cite{sathya:schutz:2009} for an ample and recent
review).

\paragraph{Transient sources}
Under this name we group all those sources emitting signals which last from
tens of ms up to a few minutes in the detector observation band; for ground
based detectors, these include for instance the coalescing binaries of low and
intermediate mass~\cite{grishchuk:2001}, and collapsing massive stars~\cite{fryer:new:2003}.

The main requirement on third-generation GW observatories is to allow the
simultaneous reconstruction of the sky location and of the source
polarization; for instance, these informations allow to estimate the source
distance in coalescing binary events, and together with an optical counterpart
this would allow a direct measurement of the Hubble constant~\cite{schutz:1986}.
The sky location in turn needs to be accurate enough to optimize the
association with electromagnetic or neutrino transients, for instance to allow
an efficient follow up in the optical band.

\paragraph{Continuous, deterministic sources}
We group here all those point sources emitting continuous wave signals, like
spinning neutron stars~\cite{bonazzola:1997}; the signals are expected to be
almost monochromatic, like those emitted by the known pulsars, but may also
display a  significant linewidth due, for instance, to the mechanism of mass
accretion, like in the low-mass X-ray binaries~\cite{ushomirsky:2000}.

In all cases, for such signals the observation time is of the order of several
months to years, and the relative motion of any individual detector and the
source allows to access also the polarization information, with an accuracy
inversely proportional to the signal to noise ratio (SNR) available.

\paragraph{Stochastic sources}
Here we consider both the stochastic background of cosmological origin, which
is generally expected to be modeled as a Gaussian, unpolarized noise, received
incoherently from all sky directions~\cite{lsc:2005}; and the background due
to a large number of astrophysical, unresolved sources~\cite{regimbau:2007}.

The noise nature of the signal requires to correlate the output of different
detectors, having uncorrelated noises, in order to estimate a background due
to gravitational waves~\cite{allen:romano:1999}; therefore, the sensitivity to
GW background scales with the product of each detector sensitivity.

\subsection{Requirements on topologies}

It seems effective to consider a few different and plausible scenarios:
a) single, ET-class observatory; b) an ET-class detector in collaboration with
several advanced detectors such as Advanced LIGO or Advanced VIRGO or
c) several ET-class detectors distributed over the world.

\subsubsection{Requirements on a single, ET-class observatory}
\label{subsub:singleET}

By an ET-class observatory we mean, as in Figure~\ref{fig:triangle} an ensemble of
instruments located at the same place on Earth, capable of reconstructing
both polarizations, with partial or full redundancy.
%; note indeed that the
%triangular one, with three interferometers, provides a single null-stream
%combination, while the two pairs of L-shaped detectors provide two independent
%null-stream combinations.

%\begin{figure}[htb]
%\begin{center}
%\includegraphics[width=0.4\textwidth,keepaspectratio,clip]{pic/AntennaPatternTriangleUncoherent.eps}
%\caption{The antenna pattern of a triangular detector, laying in the horizontal
%  plane, averaged over the source polarization.}
%\label{fig:triangleAP}
%\end{center}
%\end{figure}

\begin{figure}[t]
\begin{center}
\includegraphics[width=\textwidth,viewport=50 0 840 245,clip]{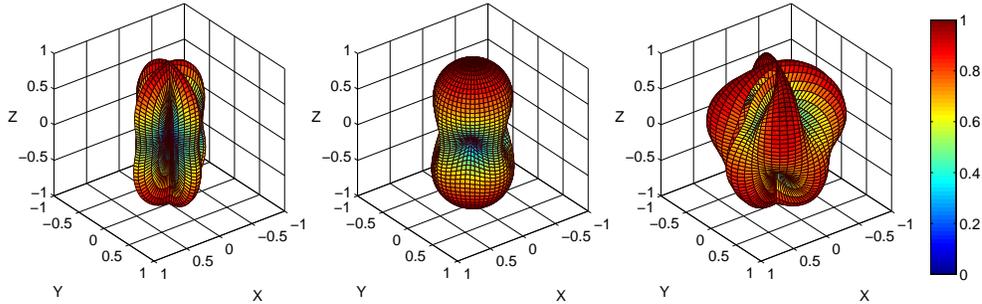}
\caption{
The response of a detector to a linear polarised gravitational wave
as a function of the detector orientation. All three plots show the normalised sensitivity
to a wave travelling along the z-axis. Each data point represents the sensitivity of the
detector for a specific detector orientation defined by the detector normal
passing the respective data point and the origin. The colour
of the data point as well as its distance from the origin indicate the magnitude of the
sensitivity. The left plot depicts the response of a single Michelson, while the center plot
gives the response of a set of three interferometers in a `flat' triangular geometry and the
right plot refers to three detectors in a pyramid geometry. %%TODO: as shown in ...
}
\label{fig:triangleAP}
\end{center}
\end{figure}

The localization of transient sources is the main weakness of a single
observatory; the antenna patterns  shown in Figure~\ref{fig:triangleAP} indicate
that all possible detector geometries feature relatively broad patterns and
offer very limited possibility of source localization.
For some sources, like coalescing binaries at a sufficient SNR,
a single observatory may still allow an approximate reconstruction of the sky
location, by exploiting the amplitude information carried by the different
harmonics of the signal; %~\cite{sathya:forthcoming} %% not allowed by Springer
however the angular accuracy
attainable is expected to be order(s) of magnitude smaller than what is
possible with a network of detectors, because of the lack of the
time-of-flight information.

For continuous, deterministic signals instead, a single ET-class detector
is sufficient to completely characterize the gravitational waves
received, and the null-stream information can be fully exploited.

For stochastic background signals, the main requirement on the topology is
that the detector antenna patterns, for each given polarization, display a
good overlap for a large fraction of sky directions.
In this case, having several detectors at the same location minimizes the loss
of signal coherency due to the finite correlation length of the stochastic
background signal, as measured by the so-called overlap reduction function
$\gamma(r f)$, where $r$ is the distance of the two detectors and $f$ represents
the signal frequency.
To provide a quantitative measure, recall that the stochastic background is
measured by the
%adimensional
quantity $\Omega_{GW}(f)$, the density of the
gravitational waves energy $\rho_{GW}$ per unit logarithmic frequency
\begin{equation}
\Omega_{GW}(f)\equiv\frac{1}{\rho_c}\frac{d\rho_{GW}}{d\log f} ,
\end{equation}
normalized to the critical energy $\rho_c$ for the Universe closure.
The sensitivity to $\Omega_{GW}$ scales as the product
$\tilde{h}_1(f)\times\tilde{h}_2(f)$ of the individual detector sensitivities;
the two Advanced LIGO detectors~\cite{advancedligo:2007} at the Hanford site
will be able to bound $\Omega_{GW}\le 10^{-9}$, and an ET detector will
improve this limit by about two orders of magnitude.

We stress again that a severe limit to the performance of co-located
instruments could result from disturbances that induce correlated noises.
The optical topology should be designed to reduce some of these effects,
for instance by requiring that different beams do not share any optical
element, or even that they are placed in different vacuum systems.
However, such solutions will decouple only some noise sources, like glitches
induced by dust crossing the beam or extra noise induced by the control system
steering the optical elements. Several other noise sources, for instance of
electromagnetic or seismic origin, including in the latter the gravity
gradient noise, will require solutions that are beyond the scope of this paper.

We should add that for continuos, narrow band gravitational waves emitted by
point sources these correlated noises are expected not to be a severe limit,
since the Doppler effect signature induced in the GW signals by the Earth
rotation and revolution should allow discriminating them.

\subsubsection{An ET-class observatory operated along with several advanced
  detectors}

In addition to the capabilities expressed by an ET-class observatory, the
coincident operation with advanced detectors, despite their inferior
sensitivity, could bring advantages.
For point sources, a signal displaying a large SNR, say about 50, in ET, might
still be above detection threshold on one or two of the advanced detectors; in
such a case, the time-of-flight measurement would allow to constrain the
source position, better than what is possible in the ET observatory alone.
We recall that, as a rule of thumb, the solid angle in which a source can be
constrained by triangulation in a three-detectors network is of the order of
\begin{equation}
\delta\theta^2 \propto \frac{c^2}{A}\delta t^2
\end{equation}
where $\delta t$ is the relative timing error, depending on the kind of
signal and the SNR, and $A$ is the area of the network.
For $A\sim 10^7$ km$^2$, and a $\delta t \sim 0.5\cdot 10^{-3}$\,s, typical for
coalescing binaries signals~\cite{acernese:2007} having SNR~$\sim 10$, this
leads to an estimated error box about 0.02\% of the solid angle,
This source localization would improve significantly the amplitude and
polarization reconstruction carried out on the basis of the signal available
in the ET observatory; however, selecting events with sufficient SNR in the
advanced detectors, 10 times less sensitive than ET, could limit the event
rate by factors as large as 1000.

For continuous sources instead the sensitivity gap cannot be filled, and
advanced detectors would be completely useless in the network.

\begin{figure}[htb]
\begin{center}
\includegraphics[width=0.6\textwidth,clip]{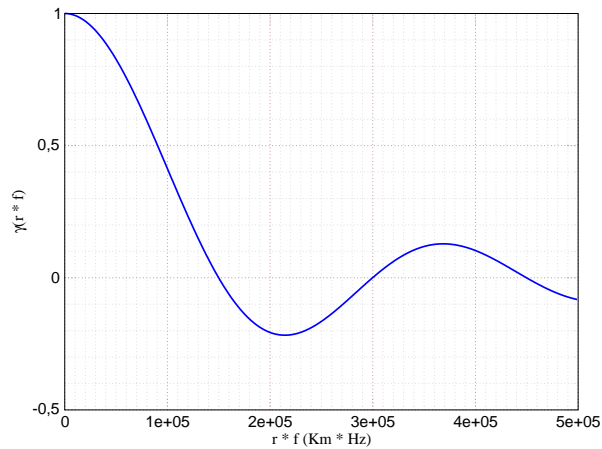}
\caption{An example of overlap reduction function, showing that detector sites
need to be closer than $\sim 300$ km in order to collect, at 300\,Hz, O(50\%)
of the signal available to co-located detectors.}
\label{fig:orf}
\end{center}
\end{figure}

For stochastic background, the correlation of data from an ET-class and an
advanced detector would lead to better upper limits by one order
of magnitude, when comparing with what is possible with a pair of not
co-located advanced detectors.
However, this estimate would be at least one order of magnitude worse than
what could be done by the ET-class observatory exploiting its own multiple
detectors. We recall, moreover, that the signal correlation will be attenuated
by the overlap reduction function $\gamma(r f)$; an example is shown in Figure~\ref{fig:orf}.

To limit the attenuation to O(50\%) at 300\,Hz, where ET displays its best
sensitivity, one should require the new detector to be built within about 300
km from one of the existing advanced detectors; this requirement, though, would
impair severely the source location capability of the mixed network, because
of the reduced effectiveness of the triangulation method.

\subsubsection{Several ET-class observatories}
This is actually not a single scenario, but could be split in two more cases:
a) each ET-class observatory hosts only one detector and
b) each ET-class observatory hosts more than one detector at the same
  location.

\paragraph{Single detector ET-class observatories}

The first case could be represented by several L-shaped detectors located
around the globe. In such a scenario, we do not have really requirements on
the optical topology, but rather we need to optimize the distance and the
relative orientation so as to maximize the scientific return.
To improve sky location reconstruction, maximizing all the distances among the
sites is the main requirement, while the relative orientation needs to find a
compromise between sky coverage and the possibility to observe the same source
with several detectors.
The distance requirement collides of course with the needs of investigations
about the stochastic background, as already discussed.
Upper limits would still be improved, with respect to non co-located advanced
detectors, by about two orders of magnitude, but could not be compared with
the sensitivity of co-located ET-class detectors.

A compromise could result by locating two of the detectors of the network at a
close distance, say up to 300km as mentioned in section~\ref{subsub:singleET},
while building the others as distant as possible on the Earth
surface. We underline that in addition to the two ``close'' detectors, we
would need at least two more ``far'' detectors for optimal source localization,

\paragraph{Multiple detector ET-class observatories}

Such a scenario could consist of two or more sites, each hosting multiple
detectors, i.e. ET-class observatories; as such, it would inherit the benefits
discussed in the previous sections, while removing most or all of the
shortcomings.
In particular, already a two-site scenario would improve dramatically the sky
localization; the time of flight measurement would allow to limit the source
position to an annulus in the sky, and additional amplitude and polarization
information could allow to further constrain it.

The measurement of the stochastic background could be carried out
independently at each site, with maximum sensitivity provided that the
correlated noise issues mentioned in section~\ref{subsub:singleET} can be
coped with; therefore the sites could be placed just as far as possible so as
to optimize their triangulation capability.
 In addition, having several sites with ET-class detectors would
allow to carry out sensitive radiometry measurements~\cite{mitra:2008},
thus mapping the background on the sky.
The large redundancy would provide, in addition to two coherent signal
channels for the gravitational wave polarizations, also a large number of
null-streams. These could be exploited, as already discussed, to reject the
noise background, but also to constrain alternative GW theories which foresee
other polarization components.
The constraints about the optical topology are derived almost directly from
what has been discussed in the previous sections, in particular it remains of
paramount importance to limit as much as possible the noise correlation among
detectors at the same site.

\begin{figure}[p]
\begin{center}
\includegraphics[width=\textwidth, viewport=0 0 680 460, clip]{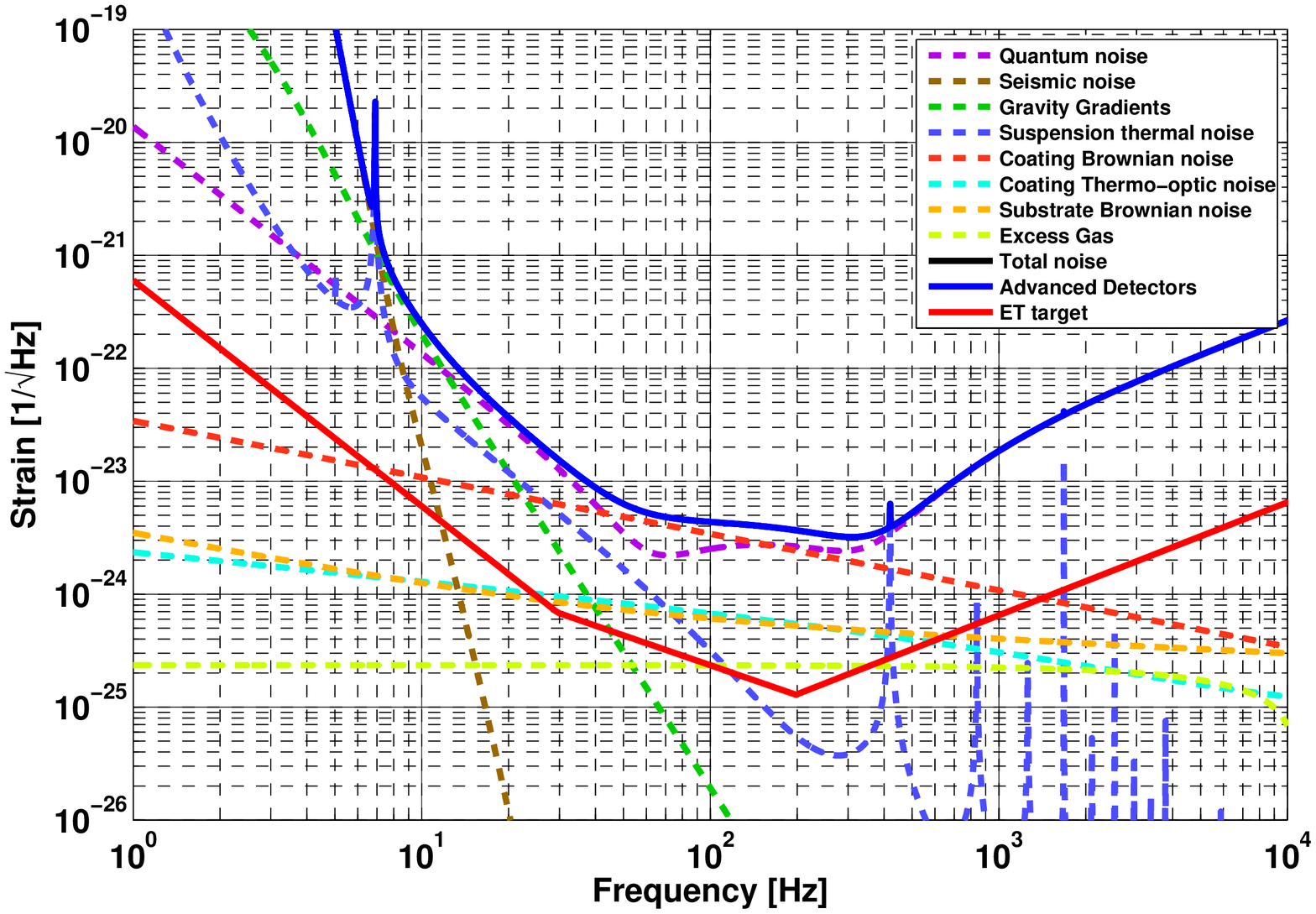}
\includegraphics[width=\textwidth, viewport=0 0 680 460, clip]{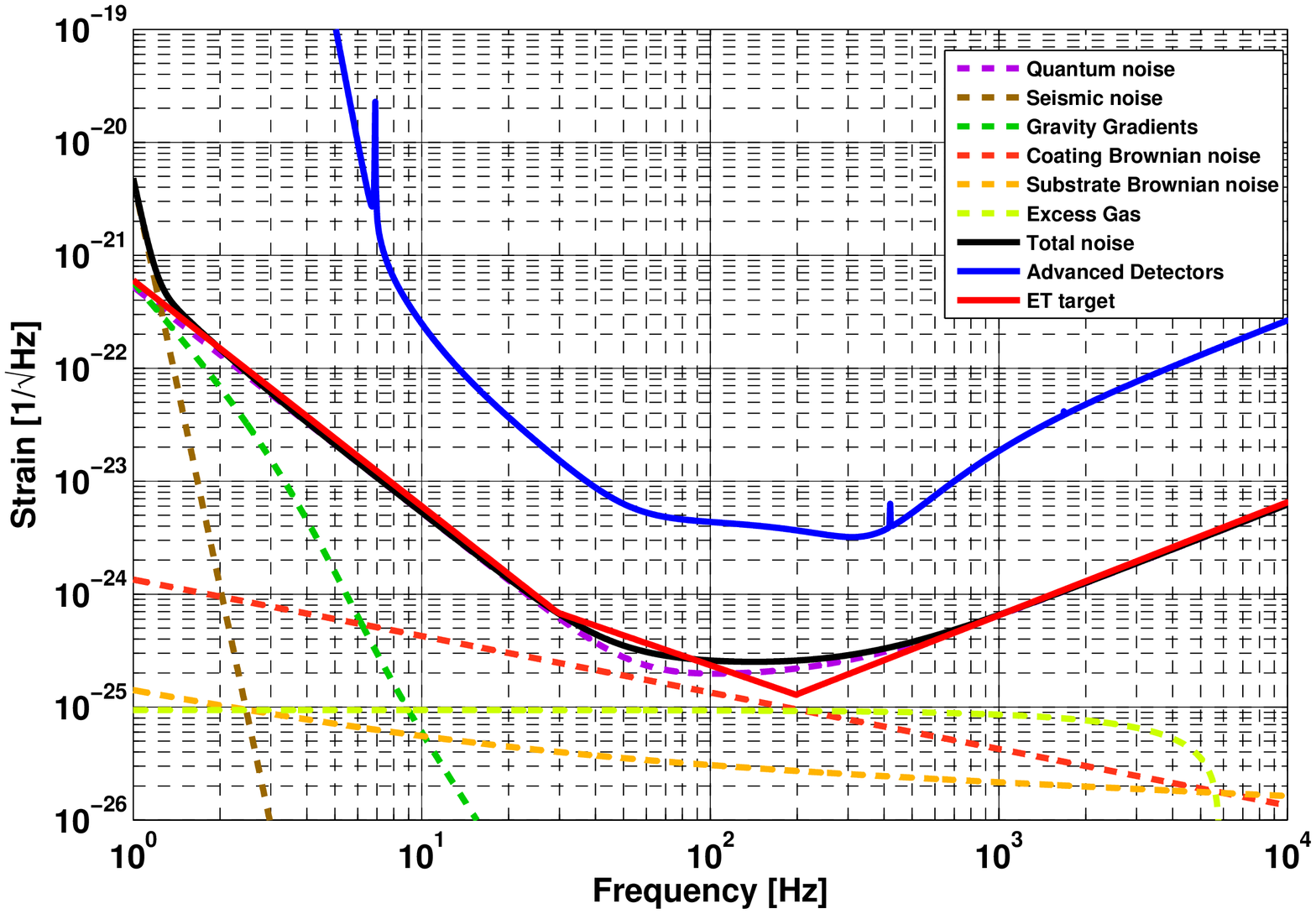}
\caption{{\bf Upper plot:} Fundamental noise contributions to the sensitivity of a potential
advanced detector (blue solid line). The solid
red line is a rough approximation of the initial ET design target.
  {\bf Lower plot:} Noise contributions and
resulting sensitivity of a potential ET configuration (black solid line) as presented in
  \cite{ET-B}. (This sensitivity option is sometimes also referred to as \emph{ET-B}
and can be downloaded from \cite{ET}, sensitivity section).
} \label{fig:noise_budget}
\end{center}
\end{figure}

\section{The brute force approach}
\label{sec:stefan}

%\tcr{
%This should extend the arXive document from Stefan, about how to reach ET
%sensitivity by extrapolating 'classical' methods. This is really interesting because
%we had many discussions but no chance to write this down properly. In effect this
%could be the core of the article.}

In this section we describe, based on \cite{ET-B},  how the ambitious sensitivity goal
for third-generation GW detectors such as the Einstein Telescope (ET)~\cite{ET}
could potentially be achieved. This approach makes use of rather conservative
technology concepts, but pushes some parameters significantly beyond what is used
in current detectors. In addition, it needs to be noted that the brute force approach
 assumes that some mayor technical problems, we currently encounter in the first generation
 GW detectors (such as for instance thermal induced mirror distortions or the abundance of
non-gaussian noise contributions) will be solved during the commissioning of the advanced
detectors.

 We start from a standard detector topology %(see Section \ref{sec:adf})
similar to a second-generation detector, i.e. a single Michelson interferometer
 with arm cavities
featuring power and signal recycling.\footnote{Several of such interferometers
could then be combined to the various detector geometries as described
in section \ref{sec:adf}.} In addition, we restrict ourselves in this section to
the application of  `conventional' technology. We use the term conventional in
 the sense that we believe the noise
reduction techniques described in this section to be a rather conservative
approach.
%, making use of the most progressed technologies currently available. 
Of course, a lot of more innovative technologies and
configurations have been suggested for application in ET.
As we will show in later sections of this article, some of these innovative
technologies, such as xylophone configurations (see section \ref{sec:ken}) and
advanced quantum noise reduction topologies (see section \ref{sec:QNR}),
might turn out to be more elegant and
less cost-intensive than the brute force approach presented in this section.
 However, as these innovative
techniques often involve a lot of new physics and have not yet been fully
tested in prototypes, we find it useful to construct a first reference design from
tested technologies.

The upper plot of Figure \ref{fig:noise_budget} shows exemplary
the individual noise contributions (dashed traces) of an advanced
detector (second generation) together with the resulting sensitivity (blue
solid trace) and a rough estimate of the targeted ET sensitivity (red
solid trace). Compared to the sensitivity of second-generation detectors
the ET sensitivity target requires a sensitivity improvement of a factor
between 10 and 50 for all frequencies above 10\,Hz. Even more
ambitious is the targeted sensitivity improvement of up to several
orders of magnitude at very low frequencies (below 10\,Hz).  It
is also worth noting that every noise contribution shown in this
plot is at least at some frequencies above the targeted ET sensitivity,
 thus, when going from second to third generation, we have to improve
every single noise contribution. One rather simple but cost intensive
way to decrease all of these noise contributions by a factor of about
3 is to increase the arm length of ET to 10\,km.  Though this helps
already a lot we have to further improve all noise contributions, as
we will discuss in more detail in the following sections.

\subsection{High-frequency noise contributions}
\label{sec:stefan:HF}
The high frequency sensitivity (above a few 100\,Hz) of advanced
detectors will be entirely
limited by photon shot noise which is the high frequency component of
quantum noise \cite{Buonanno01}. We assumed that the required shot noise
reduction of a factor of about 50, can be achieved by a combination
of the following changes. The shot noise contribution scales
inverse with the arm length of the interferometer, inverse with
the square root of the optical power stored in the arm cavities and
directly inverse with any applied quantum noise suppression, for instance from
the injection of squeezed light \cite{Caves81}. In addition, we can vary the two
signal recycling parameters \cite{Meers88} (signal recycling tuning phase and
reflectivity of the signal recycling mirror) to optimise the shape
of the quantum noise.
%Figure \ref{fig:SR} compares the quantum
%noise achieved with three different sets of signal recycling parameters
%(but otherwise identical optical properties). Due to the fact that in case of  tuned
%signal recycling (red dashed traces) both signal sidebands (upper and lower)
%can be recycled at the same time, it provides a wider detector bandwidth
% compared to detuned signal recycling (blue dashed curves).
Detuned signal recycling allows to increase the peak sensitivity,
but this comes at the cost of significantly reduced high-frequency performance.
Moreover, detuned signal recycling would make the application of
squeezed light much more hardware intensive; in this case km-long,
low-loss filter cavities would be required \cite{Kimble02}.
Therefore, it seems likely that any broad-band ET interferometer
would feature tuned signal recycling.\footnote{As we will show in
section \ref{sec:ken}, detuned signal recycling
might be an interesting option for any low-frequency interferometer of
an ET xylophone configuration.}

%\begin{figure}[htb]
%\begin{center}
%\includegraphics[width=0.7\textwidth,keepaspectratio]{./pic/SR_stuff.eps}
%\caption{\tcb{In case we have to reduce the article length, this figure could
%go out.} Quantum noise of a potential ET configuration featuring various signal
%recycling configurations combined with an assumed quantum noise reduction
%of 10\,dB. The red dashed trace refers to the configuration described in
%\cite{ET-B}, featuring tuned signal recycling (SR phase = 0) and a power
%transmittance of the signal recycling mirror (SRM tans) of 10\,\%. In comparison the
%two blue dashed curves represent potential detuned signal recycling configurations.
%%%%using the same opticla power and quantum reduction as the tuned signal recycling
%%%configuration.
%While detuned signal recycling would allow a moderate increase of sensitivity
%around the peak sensitivity, the significant bandwidth decrease, i.e. strongly
%reduced sensitivity at high frequencies, makes detuned signal recycling appear
%less beneficial for application in a broad-band ET interferometer.}
% \label{fig:SR}
%\end{center}
%\end{figure}

In summary, in \cite{ET-B} it was assumed that the shot noise contribution will be improved
firstly by a factor 2 from increasing the circulating optical power by a factor 4, from
0.75\,MW to 3\,MW, secondly by a factor of 3.3 due to the increased arm length,
thirdly by about a factor of 3 from the application of 10\,dB of broad-band
squeezing and finally by another factor of a few from changing to tuned
signal recycling.
Please note that these individual improvement factors have been chosen
in a rather arbitrary way and only indicate a single of many potential
possibilities to reach the
required shot noise reduction for ET.\footnote{For instance the experience we
gain from commissioning of the second-generation detectors will
tell us whether a further factor 4 of power increase can realistically be
achieved. In case we find any power level above 1\,MW to be impractical,
we might have to increase the targeted quantum noise suppression through
the application of squeezed light.}

\subsection{Mid-frequency noise contributions}

Apart from the already  discussed photon shot noise, for the mid-frequency
range of any potential ET observatory we have to improve the noise contribution
from residual gas pressure as well as all thermal noise contributions associated
with the test masses themselves. Overcoming the coating Brownian noise \cite{Harry07}
clearly  imposes the biggest challenge, as it needs to be reduced by about a
factor of 20 to be compliant with the targeted ET sensitivity.

%For a TEM$_{00}$ beam impinging onto a mirror of fixed reflectivity and fixed
%geometry, the coating Brownian noise contribution scales inverse with the arm length,
Generally the coating Brownian noise is inverse proportional to the beam spot size, inverse
proportional to the square root of the coating temperature and finally also depends
on the material parameters like loss angles and Young's modulus.
Improving the mechanical properties of the coating layers, i.e. searching
for alternative materials featuring lower mechanical losses, is
a field of intense ongoing research. However,
%as it is not clear that these
%efforts will yield the availability of better coating losses by the time of ET
%construction, we choose
we have considered here the conservative approach of reducing the coating
Brownian noise only by means of increasing the arm length, increasing the
spot size and cooling the test masses.

The maximal beam size on the test masses is determined on one hand by
the maximal available size of mirror bulk material and on the hand by the
achievable polishing accuracy for the surface curvature. The model described
in \cite{ET-B} assumes an increase of the beam radius by a factor of 2 from 6\,cm
to 12\,cm, which corresponds for an arm length of 10\,km to a radius
 of curvature of 5070\,m. One of the main consequences of such a large
beam size is that the mirrors have to be at least 60\,cm
in diameter to keep the clipping losses within an acceptable range.
As a nice side-effect increasing the spot size on the test masses  also
reduces substrate Brownian noise and coating thermo-optic noise and
 slightly reduces the contribution from residual gas pressure noise (due
to the larger volume of the beam).

It seems likely that the ET test masses have to be cooled to
cryogenic temperatures to make them compliant with the envisaged
sensitivity. Assuming that future research will find cryogenic coatings
with the same mechanical properties as the best currently available
room-temperature coatings, reducing the temperature of the mirrors from 290\,K
to 20\,K will further reduce coating Brownian noise by a factor of 3 to 4.
Please note that for various reasons fused silica, though standard
for room temperature interferometers, seems to be disadvantageous
at cryogenic temperatures. Sapphire and silicon are generally
considered as more likely candidates for cryogenic mirrors. An exemplary
analysis of the individual thermal noise contributions of a cryogenic
silicon test mass can be found in \cite{ET-C}.

Another way to reduce coating Brownian noise which will be discussed
in more detail in section \ref{sec:beams} is to sense the
test masses with non-conventional beam shapes.

\subsection{Low-frequency noise contributions}
\label{sec:stefan:LF}

As shown in Figure \ref{fig:noise_budget} the low frequency sensitivity
of second-generation GW observatories will be limited by a mixture of
four different fundamental noise sources: Photon radiation pressure
noise (the low-frequency component of quantum noise), seismic noise,
gravity gradient noise and suspension thermal noise. All four of these
noise contributions have to be reduced by vast amounts to reach
the ET sensitivity target.

The photon radiation pressure noise contribution scales proportional
to the square root of the circulating optical power, inverse
proportional to the arm length and inverse proportional to the mirror
mass. This means that in addition to the required improvement of a factor 20
shown in the upper graph of Figure \ref{fig:noise_budget}, we also have
to recover an additional factor of 2 originating from the factor 4 power
increase described in section \ref{sec:stefan:HF}. This total factor of
40 of required radiation pressure noise reduction can be achieved by
increasing the arm length by a factor 3.3, increasing the mirror mass
 by a factor 3 from 42 to 126\,kg,\footnote{As it was recently shown
 the coating Brownian noise of a mirror also depends
on its ratio of radius and thickness \cite{Somiya09}. It turns out that
 the mirror thickness should be roughly similar to the mirror radius
 in order to achieve reasonable coating Brownian noise. In our case with
60\,cm mirror diameter, this seems to make an even higher mirror mass
of about 200\,kg appearing more realistic \cite{ET-C}.}
 a factor 3.2 from the above mentioned
10\,dB of quantum noise suppression\footnote{A simultaneous
reduction of the shot noise and radiation pressure noise by
squeezed vacuum in a tuned RSE configuration requires one filter
cavity, see section~\ref{sec:Filtercavities} for more details.}
and a final factor of about 1.3 gained from changing to tuned
signal recycling combined with a signal recycling mirror
transmittance of 10\,\%.

\begin{table}
\begin{center}
\begin{tabular}{|l|c|c|}
\hline
& advanced detector  & potential ET design \\
\hline
Arm length & 3\,km & 10\,km \\
SR-phase & detuned (0.15\,rad) & tuned (0.0\,rad)\\
SR transmittance & 11\,\% & 10\,\% \\
Input power (after IMC) & 125\,W & 500\,W \\
Arm power & ~0.75\,MW & ~3\,MW\\
Quantum noise suppression & none & 10\,dB \\
Beam radius & ~6\,cm & 12\,cm \\
Temperature & 290\,K & 20\,K \\
Suspension & Superattenuator & 5 stages of each 10\,m length\\
Seismic (for $f>1$ \,Hz)   &  $1\cdot 10^{-7}\,{\rm m}/f^2$  (surface)  &  $5\cdot 10^{-9}\,{\rm m}/f^2$ (underground) \\
Gravity gradient reduction & none & factor 50 required \\
Mirror masses & 42\,kg & 126\,kg \\
BNS range & ~150\,Mpc & ~ 2650\,Mpc\\
BBH range & ~ 800\,Mpc & ~ 25000\,Mpc \\
\hline
\end{tabular}
\caption{Summary of the parameter changes necessary to go from an
advanced detector sensitivity to the ET design target using the approach
described in \cite{ET-B}. The second and third
columns correspond to the upper and lower graph of Figure
\ref{fig:noise_budget}, respectively.  \label{tab:summary}}
\end{center}
\end{table}

The seismic noise contribution can in general be reduced by either
reducing the ambient seismic level itself or by reducing the coupling
of the ambient seismic to the test masses, i.e. improving the
performance of the seismic isolation systems. Starting from an
ambient seismic (for frequencies above  1\,Hz)
of about $1\cdot 10^{-7}\,{\rm m}/f^2$
%(this roughly corresponds to a typical location on the earth's surface) 
we assume that the seismic level can be reduced by a factor of 20
to about $5\cdot 10^{-9}\,{\rm m}/f^2$ by building ET in an underground
location 
%(such as for instance the Kamioka mine). 
In a second, more
vigorous step we have to strongly push the seismic wall to lower
frequencies. Using passive pendulum systems this can only be achieved
by reducing the resonance frequencies of the suspensions. It has been
estimated that five pendulum stages of each 10\,m length, yielding a resonance
frequency of about 160\,mHz, together with a quiet underground
location can achieve the seismic noise suppression required for ET.

Similar to the seismic noise the gravity gradient noise contribution is
determined by the level of the ambient seismic and the corresponding
coupling transfer function to the test masses. However, the big difference
is that we have no means of influencing the 
magnitude of coupling, because we cannot shield
gravity. As shown in Figure \ref{fig:noise_budget} we need to reduce
the gravity gradient noise contribution by a challenging factor of about
3000, when progressing from second-generation detectors to
ET. We considered a factor of 3 reduction from increasing the arm length,
a factor 20 from going to an underground location (see above) and optimistically
assumed that the remaining reduction factor of 50 can be achieved
by subtracting the gravity gradient noise from the GW channel by
making use of coherent signals from an array of seismometers. For a
more comprehensive discussion of gravity gradient noise, see \cite{Cella09}.

Note that at the current stage of investigations no comprehensive
analysis of the suspension thermal noise of a cryogenic ET suspension
exists. Therefore, we omitted the suspension thermal noise trace from the
lower graph of Figure \ref{fig:noise_budget}.

\subsection{Discussion of the described sensitivity option for a single
broad-band ET observatory}

A comparison of the most important interferometer parameters
of a typical second-generation  detector and a potential ET
configuration, combining  all noise reduction efforts discussed in
this section, is shown in Table \ref{tab:summary}.
Figure \ref{fig:noise_budget} shows clearly that
the presented ET sensitivity is limited by quantum noise at
nearly all frequencies. As a consequence (and considering
the levels and slopes of gravity gradient noise and coating Brownian
noise), using heavier mirrors  would
immediately allow to improve the sensitivity between 3 and
30\,Hz significantly beyond the envisaged ET target. On the other
hand our efforts have not yet reached the originally targeted sensitivity in
the band between 100 and 300\,Hz. Improving the quantum noise in this
region by means of further increasing the optical power would simultaneously
reduce the low-frequency sensitivity. Similarly, detuned signal recycling
which offers the potential of reaching the targeted sensitivity around
200\,Hz, would worsen the sensitivity at the lower and higher frequencies.
This restriction of peak-sensitivity versus bandwidth can potentially be
overcome by so-called xylophone configurations \cite{ET-C}, i.e. by building several
narrow-band interferometers each optimised to give optimal performance
in a certain frequency band.

\section{Building a multi-band antenna}
\label{sec:ken}

As the targeted sensitivity and bandwidth for future gravitational wave detectors
is pushed
higher, eventually it must be true that it becomes better to employ multiple
interferometers rather than a single one.  There are two main reasons for this,
and both of them, as discussed below, lead to an approach by which two or
more interferometers of rather narrow bandwidth can be combined to
better deliver the required broad-band sensitivity than a single interferometer.
The first concept, the traditional multi-interferometer approach
described in section~\ref{sec:xylo:trad}, was originally proposed
to increase the high frequency sensitivity of shot noise limited interferometers
featuring signal recycling. However, more recently it was realised that
the real benefit of xylophone interferometers is the potential to
significantly improve the low frequency sensitivity by resolving incompatibilities
interferometer parameters and noise reduction schemes. This highly
profitable approach will be described in detail in section~\ref{sec:xylo:LF}.

\subsection{The traditional multi-interferometer approach}
\label{sec:xylo:trad}
The original proposal for multiple interferometers, often called a `xylophone'
by analogy to the percussion instrument, arose in the context of the shot noise
limit of relatively narrow-band signal recycled interferometers operating at high
 frequency.  In this case a classical picture taking into account only shot noise
 and coating thermal noise suffices to describe the situation and explain the approach.

The best way to characterise the performance of such a detector is by the spectral
 density of the noise and the bandwidth taken together. % (strictly the noise bandwidth
%should be considered, but the response shape is more or less fixed and a loose
%definition of bandwidth suffices for a general argument).
Here we assume %, as is very probable in 3rd generation detectors,
that the
performance limit arises mainly from shot noise, and that, at least at frequencies above
 a few $100\,$Hz, thermal noise can be neglected.\footnote{This is always true at sufficiently
 high frequency, as shot noise increases in cavity-based interferometers, and thermal
 noise decreases, with increasing frequency (ignoring resonant modes in both
 the optical and mechanical cases).}
The sensitivity is limited by shot noise because there is a limit to the optical energy
that may be stored in the interferometer.  Setting aside the technique of resonant
 sideband extraction (RSE)\footnote{RSE, a technique proposed specifically to allow
 the photon storage time to be
 increased without (to first order) affecting the signal bandwidth provides only a
 modest degree of improvement as losses in the RSE system limit the extent
 to which the storage time can be increased.}
 the storage time for the photons in the arms of the
 interferometer is capped by the desired bandwidth of the signal response.  In this
 case the energy limit is essentially a power limit -- set by the tolerable heating of
the cavity mirrors that form the arms of the interferometers.  In the first and second-generation
detectors the heating leads to unwanted and eventually uncorrectable
 distortion of the optical mode through the change of shape or change of refractive
index of the optic. In the advanced detectors, for example, it is intended to push
 the circulating power to $\sim 1\,$MW, and that requires state of the art low
absorption optical materials and aggressive thermal compensation.\footnote{In
proposed cryogenic detectors the limiting light power is not yet known, but given
 the difficulty of extracting heat at very low temperatures, it is unlikely to be much
greater than in the advanced detectors.}

\begin{figure}[htb]
\begin{center}
\includegraphics[width=0.7\textwidth,viewport=0 0 410 210, keepaspectratio]{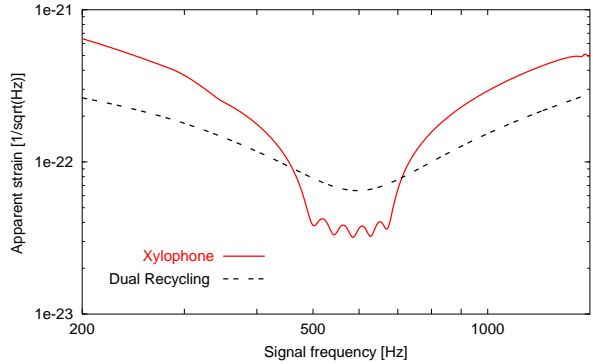}
\caption{Shot noise limited sensitivity of a xylophone interferometer  \cite{phd.Freise}
compared to an equivalent Dual-Recycled interferometer.
} \label{fig:MultipleSignalRecycledInterferometers}
\end{center}
\end{figure}

Initial xylophone ideas considered the option of spatially overlapping interferometers
sharing the same mirrors, but employing different wavelengths \cite{phd.Freise}.  Such multi-colour
detectors while potentially attractive in terms of cost, do not solve the heating problem.
Thus it is necessary to consider multiple interferometers each with their own optics,
 though possibly sharing the same site, and even vacuum envelope.  At higher
frequencies the mirrors can be relatively light as radiation pressure noise is not significant.
  As vibration isolation is also simple above $\sim 100\,$Hz the isolation and suspension
systems which otherwise constitute a significant fraction of the complexity and cost of
 a detector can be simple and compact.  Thus it is not unrealistic to consider several
 interferometers covering the band among them.
Figure~\ref{fig:MultipleSignalRecycledInterferometers} illustrates the basic approach with 5
signal recycled interferometers, combined to a xylophone detector.
As this example shows such a configuration leads to only moderately improved
sensitivity  over a limited bandwidth. Thus from the point of third generation
observatories the prospects of this traditional xylophone approach are rather limited.

\subsection{Multi-interferometer solutions for low frequencies}
\label{sec:xylo:LF}

More recently, in the development of techniques for second and third-generation detectors,
 it has been realised that multiple interferometers could become a key element
of future systems.  A more complete description of the interferometer is required
 to explain this approach.

There are several noise contributions that, broadly speaking, rise steeply
 towards low frequency, including radiation pressure noise ($f^{-2}$ in non-QND
systems), suspension thermal noise (broadly $f^{-2.5}$ except near suspension
 resonances) and gravity-gradient noise, which has a complicated spectrum.
The first two of these also scale with the mass $m$ of the mirror: directly in the
 case of radiation pressure noise and as $m^{1/4}$ in the case of suspension
 thermal noise.  In addition, seismic noise, after filtering by the isolation and
suspension systems has a much steeper gradient, but in that case it is more
convenient to represent performance by a seismic wall.  To push the wall down
to lower frequency requires larger and/or more sophisticated approaches to
isolation (see section \ref{sec:stefan:LF}).
%As an example, a single stage passive vertical isolator with a resonant
 %frequency of 2\,Hz has a sag of 6\,cm and several of these cascaded may
 %provide good isolation above $~20\,$Hz, on the other hand a similar degree
 %of isolation at 2\,Hz would require each stage to have a gravitational sag of
%6\,m.
These facts are the key ingredients in the design of xylophones for
application at low frequency (here taken to be below about 100\,Hz).

The changing mix of noise contributions as a function of frequency suggests
 that it may well be worth considering interferometers designed to work over
 a relatively narrow band -- perhaps an octave or two -- where one or two
noise contributions dominate.  The advantage in this approach comes whenever
techniques are applied to reduce two noise sources that dominate in distinct bands
 are mutually exclusive. The most prominent example for such an antagonism
is the scaling of quantum noise with the optical power. To improve the high
frequency sensitivity by means of reducing the photon shot noise contribution it is important
to increase the circulating light power. However, improving the low frequency
sensitivity by means of reducing photon radiation pressure noise requires
to decrease the optical power in the system.
Another example is the potential conflict  of very high power
 interferometers (to reduce shot noise) and cryogenics (to reduce suspension
 thermal noise), originating from residual light absorption in the mirrors and their coatings.

Such an xylophone  approach vastly increases the space of available designs, at the cost of
 system complexity.  Even with two interferometers covering the desired band
 there are many aspects that can then be individually optimised for both bands:
mirror mass, size and material; beam size and mode type;  operating temperature;
suspension type and isolation
cutoff frequency; interferometer topology including layout,
mirror transmittances, laser power, and readout scheme.  Note that many of these
 are independent so the number of combinations to be considered is large.

\begin{figure}[htb]
\begin{center}
\includegraphics[width=0.8\textwidth,keepaspectratio]{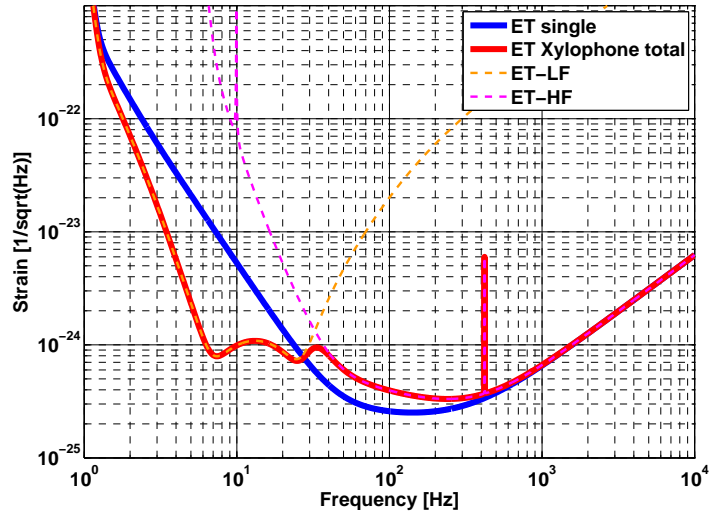}
\caption{Strain sensitivity of a potential third-generation GW
observatory as described in \cite{ET-C}. The low frequency interferometer
(ET-LF) features low optical power, cryogenic silicon optics and very tall suspensions,
while the high frequency counterpart (ET-HF) makes use of very high
optical power, silica optics at room temperature and standard suspensions.
The resulting xylophone sensitivity is compared to the single broad-band
detector described in \cite{ET-B}. (This xylophone sensitivity  is sometimes also
referred to as \emph{ET-C}
and can be downloaded from \cite{ET}, sensitivity section).
} \label{fig:ET-C}
\end{center}
\end{figure}

%As mentioned above a notable feature of the noise spectrum at low frequency
%is that it is very non-white.  This makes the optimisation process quite different
%from the high frequency case discussed in the previous Subsection.  A starting point
%that has been proposed is to consider the `facility limits' i.e.~the noise spectrum
% that would be reached with an ideal interferometer, limited only by the vacuum
%system and local gravity gradient noise (allowing perhaps  for some feed-forward
%cancellation of the latter). Unfortunately, this limit is site dependent and so not a
% good basis for general discussion, a few general trends can, however, be identified.
%
%Suspension thermal noise is expected to have a spectral shape of broadly
% $f^{-2.5}$  in displacement for most of the materials from which the suspension
% could be built.  The noise depends on the mass suspended in a complicated
%way, as there are several dissipation mechanisms involved. For example, in the
% current fused silica technology the predominant dissipation arises from the
% thermoelastic effect which has a strong frequency dependence, whereas at lower
% frequency and/or in other materials the internal structural damping can dominate.
%Likewise the dependence of suspension thermal noise on temperature is complicated
% as different mechanisms dominate at different temperatures.

One approach that has been proposed is to keep the interferometric sensing as simple
 as possible in the low frequency band, i.e.~without the use of a QND measurement.
 In that case the balance between shot noise and radiation pressure is simple and indeed,
 as the point is to use low light power in the lowest frequency bands, only shot noise,
 with a white spectrum, need be considered.  This approach leads to a cascade of
 interferometers with massive mirrors, cryogenic suspensions, aggressive vibration
 isolation, and modest light power at the low end, while at the high end the mirrors may
 be smaller, the suspensions need not be cryogenic, vibration isolation is simply
 achieved, but the light power must be high (perhaps also with squeezing or QND
 techniques). The bands would possibly be approximately logarithmic each covering
 somewhere between an octave and a decade in frequency.

A different approach has been taken in work towards the design of the Einstein
Telescope \cite{ET-C}.  In this case the idea was to split the whole detection band, from below
 10\,Hz to above 1\,kHz in two. The two interferometers both employ mirrors
weighing about 200\,kg, with the interferometers basically
configured in the traditional dual recycled Fabry-Perot Michelson
proposed for the second-generation detectors although enhanced by
including systems to suppress quantum noise by 10\,dB in both
cases\footnote{To achieve a broadband quantum noise suppression by
e.g. a factor of 10\,dB using squeezed vacuum, the low frequency
detector requires two filter cavities due to its detuned RSE
configuration, whereas the high frequency detector requires only
one filter cavity, see section~\ref{sec:Filtercavities} for more
details.}, but while the high frequency (HF) detector uses fused
silica at room temperature,
 the LF detector mirrors are single crystal silicon for good performance at low
temperature (10\,K).  The low frequency (LF) detector has 50\,m tall suspensions while those
of the HF detector are a few meters tall; the interferometry for the LF system uses
just 18\,kW of circulating power, for low heating with a conventional Gaussian beam
shape and has a response intended to give maximum sensitivity from about
 $7\sim30\,$Hz,  while the HF interferometer has 3\,MW of circulating power and employs
 Laguerre-Gaussian 3,3 mode beams to better average over thermal noise
 in the mirror coatings.  Another difference is that the LF interferometer uses an
optical wavelength of 1550\,nm at which silicon is transparent, unlike the 1064\,nm
 suited to the silica mirrors of the HF interferometer.  The strain sensitivities
of the LF and HF interferometer as well as the resulting xylophone
sensitivity are shown in Figure~\ref{fig:ET-C}. Although this system is
 proposed as a straw-man and the detail is subject to change, it provides a
good illustration of the complexity and variety of choice of parameters for
 even a two band detector.

%%%%%%%%%%%%%%%%%%%%%%%%%%%%%%%%%%%%%%%
%
% Newly added on July 23rd by Kentaro
%
%%%%%%%%%%%%%%%%%%%%%%%%%%%%%%%%%%%%%%%

\section{Quantum noise reduction}
\label{sec:QNR}

We have shown that the target sensitivity of a detector as envisaged in the
ET design study could potentially be achieved by an
interferometer in a {\it conventional} configuration with ultimately
good parameters that we can expect in the future. The xylophone option
will ease the use of some incompatible parameters. On the other hand,
we also know that there are number of configurations theoretically
developed to be used for a future detector. These {\it advanced}
configurations would let us achieve the same sensitivity increase with easier
parameters.
%%% adf 310110, fixing some English problems below.
%Besides, it is not to be blamed to realize even better sensitivity.
Most of what we shall show in the following sections has not been
experimentally demonstrated but is expected to be in time for the
construction of third-generation detectors. In this section, we
focus on the reduction of quantum noise. At first we present in
Sec.~\ref{sec:Filtercavities} how filter cavities can be used to
transform a conventional interferometer using squeezed vacuum into
either a \textit{frequency dependent squeezed-input
interferometer} or a \textit{variational-output interferometer} to
reduce the quantum. In Sec.~\ref{sec:speedmeter} we will
introduce the concept of the {\it speedmeter} which can help
reducing the quantum radiation-pressure noise whereas in
Sec.~\ref{sec:opticalbar}, we introduce a {\it optical-bar}
regime, which is another concept for reducing quantum noise. More detailed
explanation of these scheme is given in a separate article in this
issue by Danilishin {\it et al}~\cite{GRG-QND}. Further
information can be found in a direct comparison of several quantum
noise reduction schemes which has been performed as part of the ET
design study~\cite{et_mueller_ebhard09}. In
Sec.~\ref{sec:kentaro}, we will introduce some more advanced
configurations for the reduction of classical noise.

\subsection{Filter cavity}\label{sec:Filtercavities}
\begin{figure}[t]
\begin{center}
\includegraphics*[width=1\textwidth]{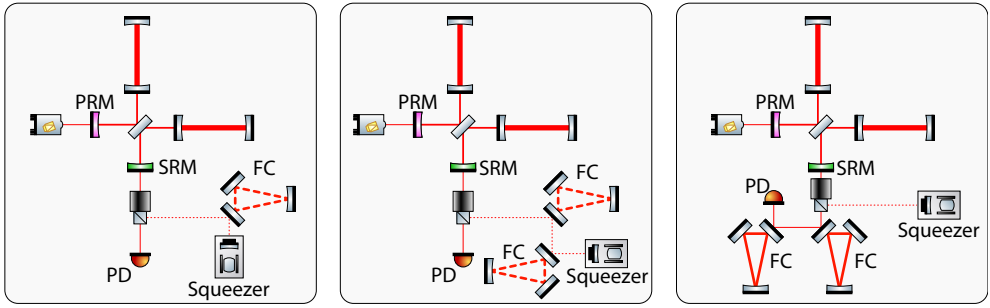}
\caption{{\it Left panel}: Tuned RSE interferometer using one FC
to create a frequency dependent squeezed input achieving a
broadband quantum noise reduction. {\it Middle panel}: Detuned RSE
interferometer using two FCs to create a frequency dependent
squeezed input achieving a broadband quantum noise reduction. {\it
Right panel}: RSE interferometer using a pair of filter cavities
thereby changing the readout configuration to variational output.
FC: filter cavity; PD: photo diode; PRM: power-recycling mirror;
SRM: signal-recycling mirror.}
   \label{fig:RSE-with-FC}
\end{center}
\end{figure}
In general, filter cavities can be used to turn a conventional
interferometer either into a \textit{frequency dependent
squeezed-input interferometer} where an initial frequency
independent squeezed vacuum is made frequency dependent before it
is injected into the interferometer or into a
\textit{variational-output interferometer} where the readout
quadrature of output light is made frequency dependent to maximise
the SNR of the detector. The performance of these two
approaches is different. Where the frequency dependent
squeezed-input interferometer achieves maximally a broadband
reduction of the initial quantum noise by the squeezing strength,
the variational-output interferometer additionally allows a
complete cancelation of the radiation pressure noise so that only
the shot noise component of the quantum noise remains in the
signal output. Layouts of both schemes for the RSE topology are
shown in Figure~\ref{fig:RSE-with-FC}. For more information about
frequency dependent squeezed-input and variational-output the
reader is referred to \cite{Kimble02,Harms03,Buonanno04}. In both
cases the light is reflected at one or several filter cavities to
imprint a certain frequency dependent phase shift. The
configuration of the interferometer determines hereby the number
of filter cavities required as well as their parameters to achieve
a certain quantum noise reduction. In the case of a single band
antenna approach as presented in section~\ref{sec:stefan} a tuned
RSE configuration is used. As a consequence a direct injection of
phase quadrature squeezed vacuum achieves a broadband reduction of
the quantum noise only in the shot noise limited regime. To
achieve a full broadband reduction in the whole frequency range of
the detector, a single filter cavity is necessary to transform the
initial phase squeezed vacuum into amplitude squeezed
vacuum to reduce the quantum noise in the radiation pressure
dominated frequency range simultaneously to the shot noise
dominated regime. The same topology but using a detuned RSE
configuration, is used in the low frequency detector of the
multi-band antenna approach in section~\ref{sec:ken}. It requires
two filter cavities to achieve the same broadband improvement in
the whole frequency range.
%
%Here the second filter cavity is used to
Both filter cavities together
compensate the dispersion introduced by the detuned signal
recycling cavity and the resulting non-symmetric phase delay
applied to the two signal sidebands of sideband frequencies
$\pm\Omega$ with respect to the carrier at frequency $\omega_0$.
%Table~\ref{tab:fitercavity} lists the parameters of an example
%filter cavity that compensates the effects of the detuned signal
%recycling cavity.
In both RSE configuration cases, tuned or detuned, and assuming no
losses the initial quantum noise curve can be shifted downwards at
all frequencies by the squeezing factor as stated in
Table~\ref{tab:summary} resulting in the new squeezed quantum
noise. The concept of the filter cavity has already been
demonstrated in a table top experiment in 2005 \cite{Chelkowski05}
shortly followed by the demonstration of a broadband improvement
of the quantum noise by squeezed vacuum in a detuned signal
recycled configuration \cite{Vahlbruch05}.

%\begin{table}
%\begin{center}
%\begin{tabular}{|l|c|}
%\hline
%& Filter cavity parameters\\
%\hline
%geometrical length & 1\,km\\
%Finesse & 596\\
%Transmissivity  of the input/output coupler & $\approx$\,5250\,ppm\\
%Transmissivity of the curved HR mirror & 30\,ppm\\
%\hline
%\end{tabular}
%\caption{Example of parameters for a triangular filter cavity to
%compensate the effects of a detuned signal recycling cavity in the
%single band antenna approach presented in
%Section~\ref{sec:stefan}.} \label{tab:fitercavity}
%\end{center}
%\end{table}

\subsection{Speedmeter}\label{sec:speedmeter}

%There are several attempts to realize this cancellation in a broad frequency band,
% which is called {\it variational readout}. The only way so far to realize the idea
% is with a so-called {\it filter cavity}
%(see the right panel of Figure~\ref{fig:pondero})
%introduced by Kimble {\it et al}~\cite{Kimble02}. The output field of the interferometer
%is injected to a pair of additional cavities to add some phase shifts depending on the
% signal frequency so that the optimal quadrature to cancel radiation pressure noise
% is fixed in a broad frequency band.
%
%In fact, the variational readout is highly sensitive to the optical
%losses, i.e. absorption and scattering at a mirror and quantum
%inefficiency. The loss vacuum is incoherent to the ponderomotively
%squeezed vacuum and can be larger than that especially in the optimal
%quadrature.

A so-called {\it speedmeter} is another way to suppress radiation
pressure noise in broadband. There are several ways to realize a
speedmeter, one of which is the well-known Sagnac configuration as is
shown in the left panel of Figure~\ref{fig:Sag-OB}. The incident light
is split into two and probes the motion of the mirrors in each arm
with a certain time delay. The vacuum field from the signal extraction
port is split into two and imposes radiation pressure noise with the
time delay. Thus the $a_1(\Omega)$ component is partially cancelled
($\Omega$ is multiplied). One difference is that, unlike the
variational readout, this cancellation is made before the
photo-detection, so the quantum inefficiency, supposedly the largest
part of the optical losses, is not a big problem for the
speedmeter. Although the optical loss of the mirrors limits the
ability of the speedmeter, the estimated noise curves with reasonable
amount of losses for ET, the speedmeter turns out to be the best
candidate for the low frequency measurement~\cite{GRG-QND}.

\subsection{Optical bars}\label{sec:opticalbar}

\begin{figure}[t]
\begin{center}
 \includegraphics*[width=.5\textwidth]{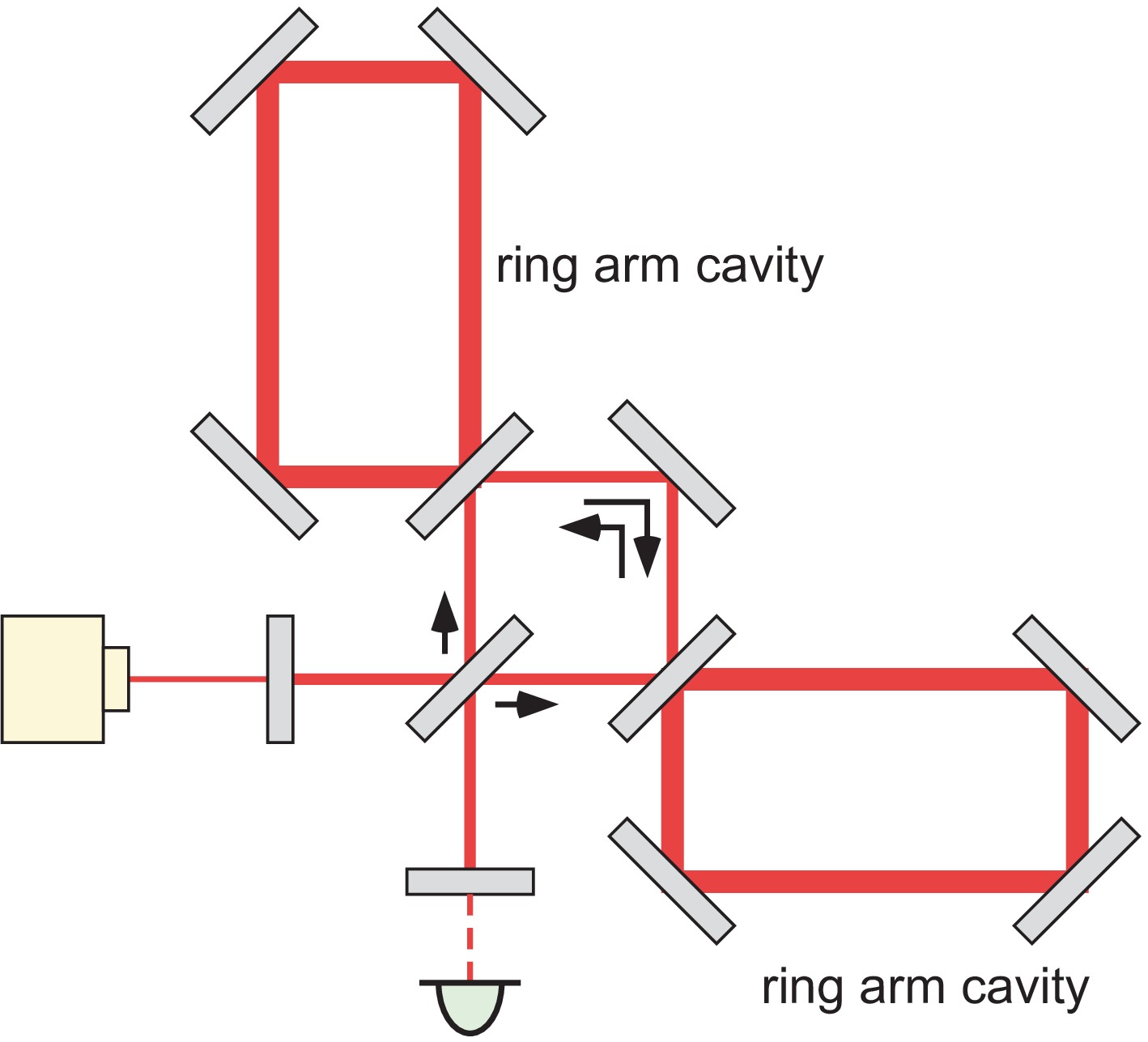}
 \hfill
 \includegraphics*[width=.4\textwidth]{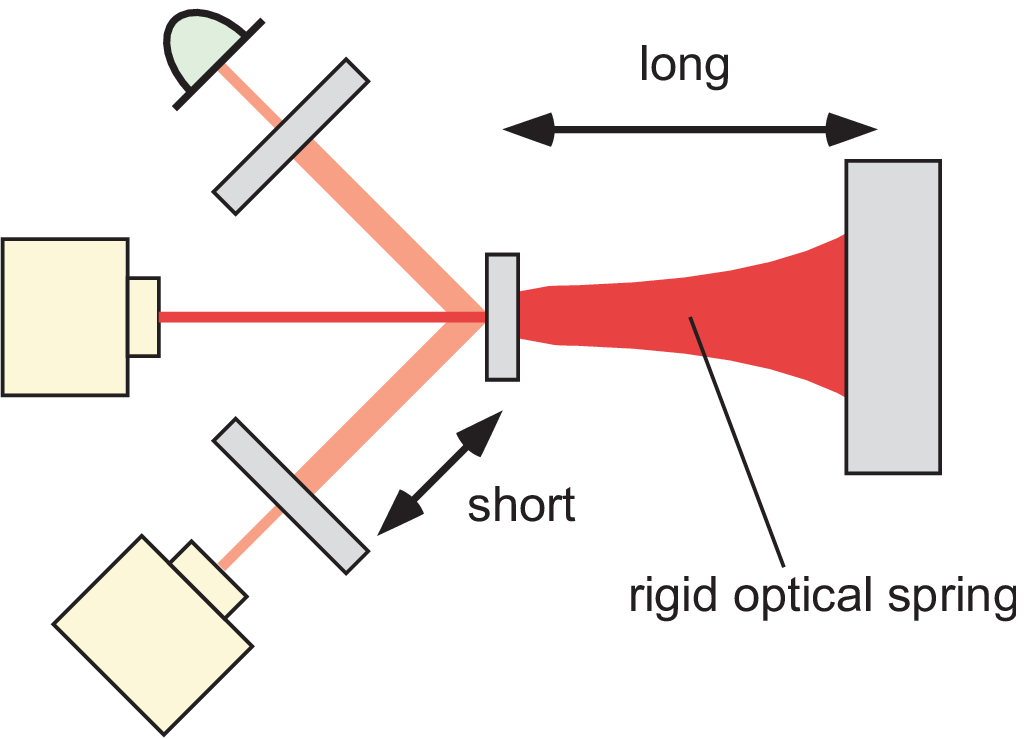}
 \caption{{\it Left panel}: Resonant Sagnac interferometer. {\it Right panel}: Optical bar.\label{fig:Sag-OB}}
\end{center}
\end{figure}

The sensitivity of ET at high frequencies is limited by shot
noise. The noise level can be reduced by injecting high power beam which
causes various problems related to heating optical components through
light absorption. Injecting squeezed vacuum from
the signal extraction port helps reducing shot noise, but it is
limited by the optical losses. Signal recycling tuned at high
frequencies can reduce shot noise but it increases the optical losses
and ruins the effect of squeezing. After all it is not easy to improve
the sensitivity at high frequencies by conventional means. One way to circumvent the problem
is to use a so-called {\it optical-bar} scheme (right panel,
Figure~\ref{fig:Sag-OB}). In this scheme, a long arm cavity is locked
by an optical spring with the resonant frequency higher than the
observation band; the masses should be small and the circulating power
should be high. The motion of the rigidly locked cavity is measured by
another interferometer that is much shorter than the locked arm
cavity.

The arm cavity is as long as a conventional detector so that the strain caused
 by gravitational waves is the
same, while the sensing part is short so that the cavity pole is pushed to a higher frequency.

A detuned signal-recycled interferometer, which is the configuration
of the second-generation gravitational-wave detectors, has the feature
of the optical bar scheme. In fact, the optical spring frequency is
close to the lower end of the observation band, so the signal
enhancement at the spring frequency is the main purpose to employ the
detuned configuration. It is, however, pointed out that the
sensitivity can be improved by an auxiliary probe light to measure the
mirror motion imposed by gravitational waves at frequencies lower than
the optical-spring resonance; this is called a {\it local readout
  scheme}~\cite{LocalReadout}.

%%%%%%%%%%%%%%%%%%%%%%%%%%%%%%%%%%%%%%%%%

%
\section{Reducing thermal effects with exotic beam shapes}\label{sec:beams}

As stated earlier overcoming the coating Brownian thermal noise
\cite{Gorodetsky08} imposes one of the biggest challenges for
third generation gravitational wave detectors. This section reviews another
interferometric concept (which thus has some impact on the topology and
configuration of future detectors) which can be used to reduce said noise.
The Brownian thermal noise in general is a consequence of the non-zero thermal
energy causing random motions of e.g. the mirrors' reflecting
surface \cite{Vinet09}. Currently there exist several techniques
to reduce the thermal noise content sensed by a gravitational wave
detector:
\begin{itemize}
  \item {Cooling of mirror test masses:} This technique directly
  reduces the thermal energy $k_BT $ stored in the mirror test mass substrates of
  the interferometer.
%Hence, cooling the mirrors to liquid helium
 % temperatures reduces the Brownian thermal noise as well as the
 % thermoelastic noise dramatically. Furthermore
  A clever choice of
  the substrate material %to one which has a non existing thermal expansion coefficient at
  %cryogenic temperatures
  allows the simultaneous reduction of Brownian thermal noise and  thermo-elastic noise \cite{Rowan05}.
  \item Use of {high-reflective mirrors with less coating layers:}
  Several research efforts are under way to reduce the coating Brownian noise
  by essentially reducing the coating thickness needed for creating the required
  high reflectivity. One such technique, the \emph{end-mirror cavity} is briefly
  explained in section~\ref{sec:endcavity}. Another approach
  is to use mirrors with a single  waveguide coating layer to achieve a high reflectivity \cite{Gossler07}.
  Both techniques are in the early prototype stages but are considered very promising.
  \item {Change the mode shape of the laser:} A change of
  the laser beams mode shape used in gravitational wave detectors
  can lower several thermal noise contributions simultaneously. \cite{Vinet09}.
\end{itemize}
This section focuses on the method using alternative beam shapes.
  While this technique is relatively new, its technical readiness is more advanced such that
  is currently considered for potential upgrades of advanced detectors.

\subsection{Prospects of alternative beam shapes}

Various alternative beam shapes have been proposed to be used in
gravitational wave detectors to reduce the measured thermal noise
as well as to lower thermal effects in general. In the following
we will concentrate on two beam shapes;
%flat beams \cite{DAmbrosio03},
the so-called Mesa beams \cite{DAmbrosio04} and higher
order Laguerre-Gauss (LG) modes \cite{Vinet05}. In these beam shapes
that laser power is distributed more widely compared to the
currently used TEM$_{00}$ mode. Hence, the fluctuations e.g.
induced by the Brownian motion of the mirrors reflective surface
are canceled significantly by averaging over a larger region of
the mirror with the cross section of the beam used.
%To make a fair
%comparison of the performance of the different mode shapes the
%clipping losses that each mode experiences at a mirror inside the
%interferometer should be constant. This ensures that effects
%induced by stray light and others related to the light power used
%in the interferometer stay the same.

At first glance a perfectly flat beam would appear to be the best candidate in terms
of thermal noise for an alternative beam shape, because of its
uniformly distributed intensity profile. But such a rectangular
shaped beam intensity will spread out very quickly while
propagating, due to diffraction. This renders flat beams
impractical to be used in the long interferometer arms of a
gravitational wave detector. To circumvent this problem the Mesa
beams have been introduced. Their intensity profile has much
smoother edges compared to the flat beam thereby reducing the beam
spread to an acceptable level. The phase fronts of these Mesa
beams require mirrors that have a \textit{Mexican Hat} profile.
The fabrication of these mirrors is more complex compared to the
currently used spherical mirrors but first test experiment with a
mesa beam based cavity has been performed at Caltech
\cite{agresti06b}. %\tcr{could not find a newer paper of the experiment}
However, the use of non-spherical mirrors is new and much more
work is required to reach the level of accuracy which can currently be
achieved in interferometers with spherical mirrors. Therefor the
use of higher order Laguerre Gauss modes has been proposed.
While LG modes can also feature a wide intensity  distribution
their phase fronts are spherical and they can be used with spherical
mirrors.. %like the currently used TEM$_{00}$ mode.

An excellent overview about the expected thermal noise of all the
alternative beam shapes mentioned above is given by Vinet
\cite{Vinet09}. %Using his equations one can e.g. calculate the
%coating Brownian thermal noise for the different mode shapes where
%each mode experiences the same 1\,ppm clipping loss at an
%interferometer mirror. Comparing the results one
We find that the
linear spectral density of the coating Brownian thermal noise of
the mesa beam is a factor of $\sim\!1.5$ lower compared to the
TEM$_{00}$ mode. The LG$_{33}$ mode and LG$_{55}$
perform even better with reduction factors of $\sim\!2.2$ and
$\sim\!2.3$ respectively. For the substrate thermal noise the
results are similar; both LG modes outperform the Mesa beam
\cite{Mours06}. Nevertheless, the calculation of the thermo-elastic
noise for the same 1\,ppm clipping loss modes only shows a
performance increase for the Mesa beam with a reduction factor of
$\sim\!1.8$ compared to the TEM$_{00}$ mode. The two LG modes
produce a higher thermoelastic noise which is larger by a factor
of $\sim\!1.7$ in case of the LG$_{33}$ mode and larger by a
factor of $\sim\!2.5$ in case of the LG$_{55}$ mode with respect
to the TEM$_{00}$ mode. In the current ET scenario the
thermo-elastic noise is far away from being the limiting thermal
noise, however, this relies on the parameters used and in general a careful trade-off in
terms of beam size and clipping loss has to be made for optimising the thermal noise
contribution using LG modes.

\subsection{Impact of alternative beam shapes on detector configuration}

LG modes are expected to be fully compatible with the noise requirements
of high-precision measurements done by gravitational wave detectors~\cite{Chelkowski09}.
Experimental verification of such theoretical studies are in
progress. Of particular interest is the impact of the mode degeneracy in
optical cavities which are resonant for a higher-order Gaussian mode.

For any alternative beam shape considered, the  focusing telescopes
and all the input/output optics component should be adapted to the new
beam shape. Also, the compatibility between alternative beam shape and
squeezing techniques should be studied.

As already pointed out, the use of Mesa beam requires particular
non-spherical mirrors, and create matching losses between the
input-output beams and the cavity beam. An complete analysis of a
\textit{mesa beam interferometer }  (i.e. with power and signal
recycling) has not been performed.

Higher-order helical Laguerre-Gaussian beams are
not compatible with the use of triangular cavities, largely used in
interferometric detectors as pre-mode cleaners, pre-stabilization
cavities, input and output mode-cleaner. The reason is the 180 degrees
phase shift for the field distribution in the plane of the cavity due
to three mirror reflections. Mode-cleaner cavities with an even number
of mirrors should be used, or in alternatively sinusoidal
Laguerre-Gauss modes.

\subsection{Generation of alternative beam shapes}

One of the challenges in using alternative beam shapes is their
generation at high power and with low amplitude- and frequency noise
as well as the required stability in beam shape and position. Typically we
expect to be able to use laser amplifiers and mode-cleaning
cavities similarly as for the TEM$_{00}$ beams. However, dedicated
research for created such light sources for gravitational wave
detectors has just begun.

One well-known method of converting a Gaussian beam into an
alternative beam shape is to use an optical cavity which is resonant only (or
dominantly) for the required beam shape.
%The mesa beam can be produced injecting a pure gaussian mode in a
%cavity formed with non-spherical \textit{mexican hat}
%mirrors.
Experiments using this method in the context of
Mesa beams have been performed in Stanford
\cite{beyesdorf2006} and Caltech  \cite{tarallo2007}.
%The price to pay
%to have an an input beam different from the stored one are
%mode-matching losses, but
If the parameters of the Gaussian beam are
optimized to have the highest overlap with the Mesa beam, the power
coupled can theoretically be more than 90\%~\cite{DAmbrosio03}. The
matching  can be further increased to shaping the input beam, for
example using deformable mirrors~\cite{avino2006}.

Higher-order Laguerre-Gauss modes have been widely studied for their
property to carry orbital angular momentum~\cite{allen92}, with
several applications in cold atom physics, quantum physics and quantum
communications~\cite{heckenberg,mair}.  Because of this
applications, most of their production techniques were mostly focused
on doughnut-like mode. Among
these techniques there are diffractive optics, phase plates coupled
with astigmatic mode converters, computer generated holograms and
optical fibers coupled with long period gratings.
For example, LG$_{21}$ modes have been produced using two diffractive optical
elements (DOE), starting from a pure Gaussian beam, with a conversion
efficiency of the order of 60\%~\cite{kennedy2002}. An incoming
Gaussian mode diffracts on the micro-structures etched on the DOEs and
the Laguerre-Gauss mode results from the sum of the diffracted
waves. The limitation on the conversion efficiency and on the mode
purity comes by the discreteness of the micro-structures. This
technique is commonly used to shape beams for industrial applications,
but the goal in that case is only to shape the power profile of the
beam and not to control the wavefront.

Another common technique to generate LG modes is to use an astigmatic
mode converter~\cite{beijersbergen1993}. The principle of an astigmatic mode converter relies on the Gouy phase
shift introduced by a pair of cylindrical lenses separated by a
suitable distance, which is a function of the focal length.
This setup transforms, in principle without any losses, an Hermite-Gaussian mode in a Laguerre-Gaussian mode. The
Hermite-Gaussian mode can be obtained passing a Gaussian beam through
a phase plate, or forcing the laser cavity to emit directly the HG
using a spatial filter. The potential drawback of the astigmatic
converter is due to the high order of the Hermite-Gaussian mode
needed: for example, to create a LG$_{33}$,  a HG$_{63}$ is needed. A
doughnut-like (LG$_{10}$) mode has been recently created using this
technique,  with a purity higher than 99\%~\cite{chu2008}; in this
case the Hermite-Gauss was created inside the laser cavity, by use of
a phase plate.

Computer generated holograms, obtained by computing the
theoretical interference between a Laguerre-Gauss mode and a plane
wave, have been used mostly to produce doughnut-like
modes, but the production of multi-ringed modes have been also
demonstrated \cite{arlt1998}. It should be remarked that this
technique does not produce pure Laguerre-Gauss modes, but a
superposition of LG modes with different quantum numbers, and this can
limit the conversion efficiency.

All the methods of
producing LG modes for quantum optics application are promising but
past research rarely include the modes of interest for GW detection, not did it investigate the excess noise
of the new light source regarding amplitude, frequency and beam jitter.
In general, the purity of the Laguerre-Gauss mode can be
increased injecting the beam in a mode-cleaner Fabry-Perot cavity. If
the cavity is kept resonant on the desired Laguerre-Gauss mode, the
unwanted spatial impurities appears under the form of non-resonant
optical modes, and they are filtered by the the cavity, proportionally
to its finesse. Furthermore, because the very high power needed in
future gravitational wave interferometers, the compatibility with high
power of the mode transformer should be addressed.

%%%%%%%%%%%%%%%%%%%%%%%%%%%%%%%%%%%%%%%%%%%%
%%%%%% Newly added on June 16th (Kentaro)
%%%%%% and changed on July 23rd (Kentaro)
%%%%%%%%%%%%%%%%%%%%%%%%%%%%%%%%%%%%%%%%%%%%

\section{Further techniques to reduce classical noise}
\label{sec:kentaro}

In this section, we introduce two examples of other advanced configurations which
have been proposed for reducing classical noise. Both of them will work as an auxiliary system
that can be added to the basic interferometer. We can regard these techniques as alternatives to realize the ET sensitivity with easier
parameters or to upgrade the sensitivity.

\subsection{Suspension point interferometry}

The ET design study envisages to be built a detector deep underground in a hard-rock mountain so that both
seismic noise and gravity-gradient noise are one or two orders smaller than the
 first or second-generation gravitational-wave detectors. The mirrors will be
suspended by multiple pendulums, each of which can be 10\,m tall.
%the resonant frequency of the pendulum is 160~mHz.

\begin{figure}[t]
\begin{center}
 \includegraphics*[width=.35\textwidth]{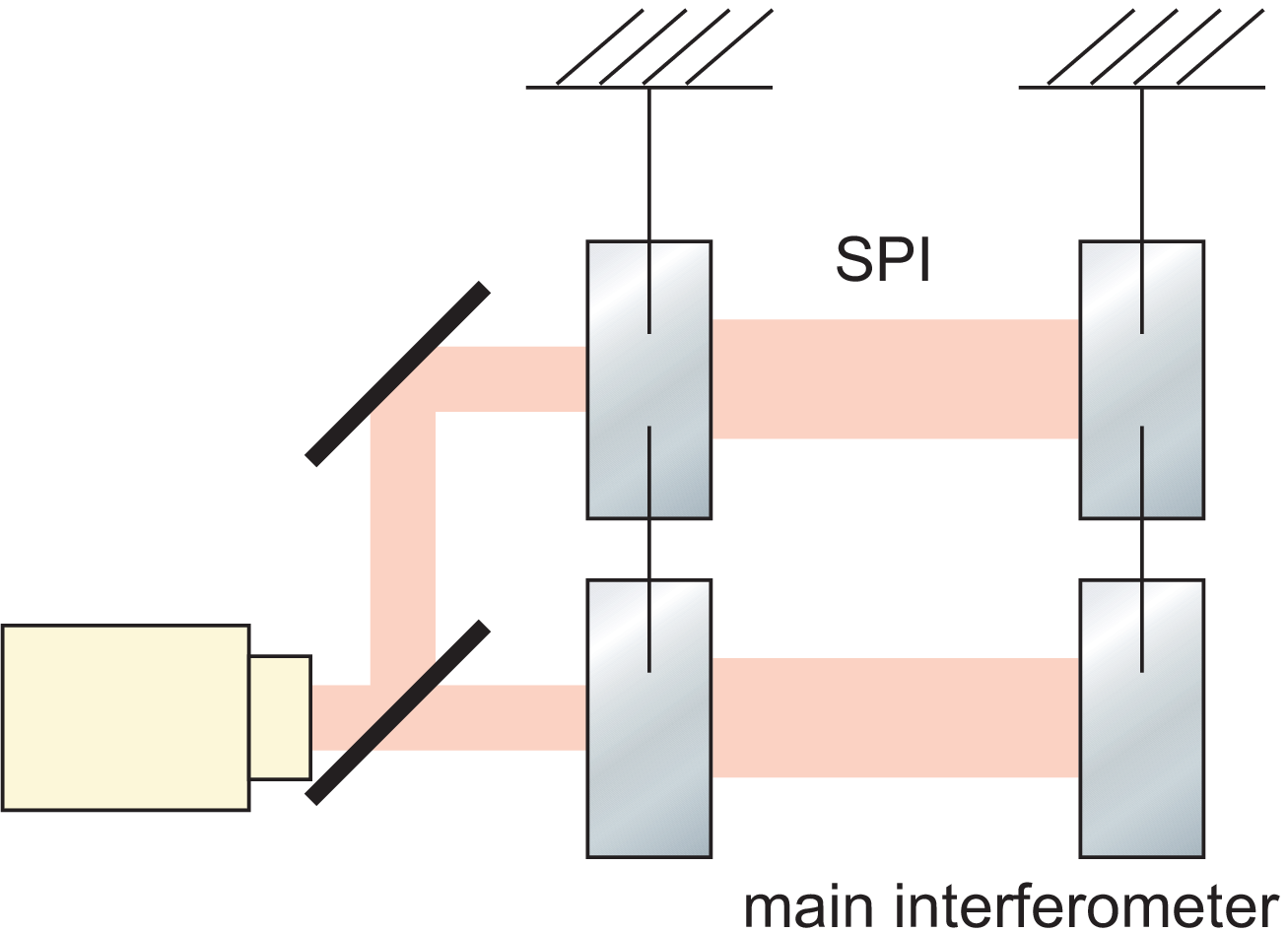}
\hfill
 \includegraphics*[width=.6\textwidth]{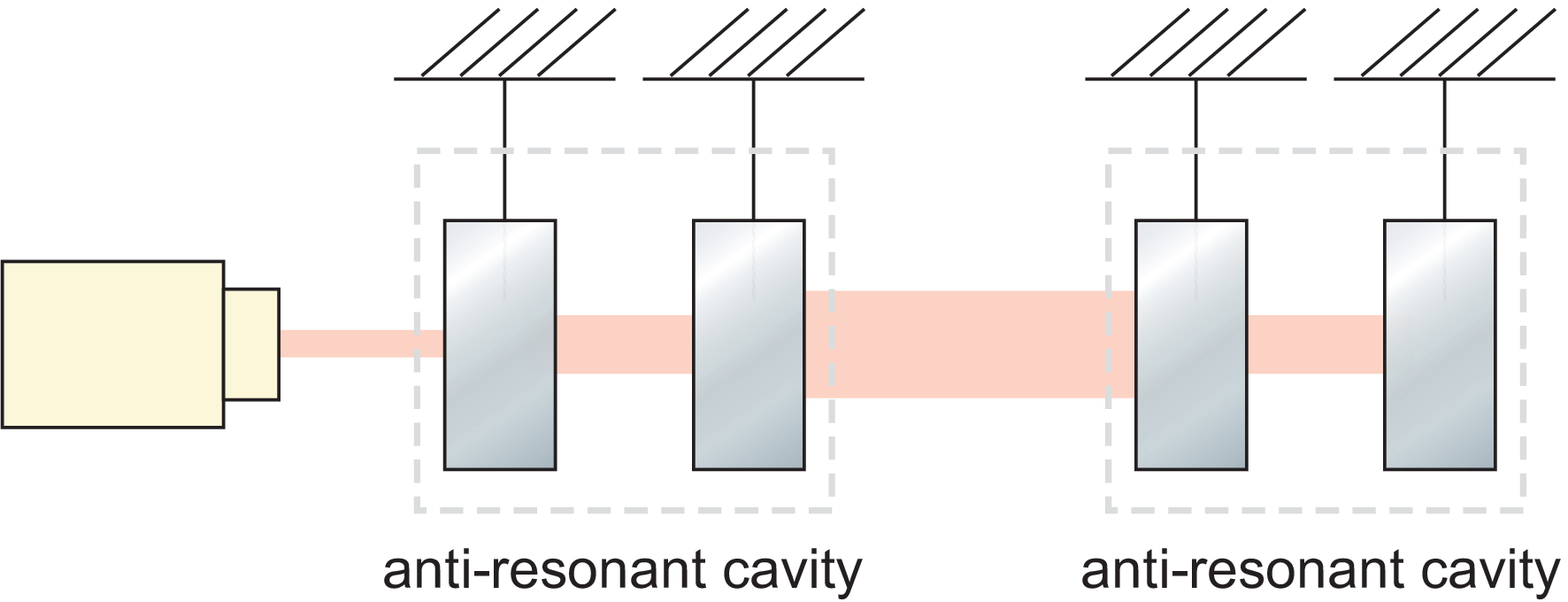}
 \caption{{\it Left panel}: The suspension-point interferometer cancels out seismic noise by a factor depending on the imbalance of the suspensions. {\it Right panel}: The end-mirror cavity (and the front-mirror cavity) realizes the mechanical separation of coating layers and also let us control and cancel out a big fraction of coating thermal noise.\label{fig:SPI-ETMC}}
\end{center}
\end{figure}

Further improvement can be made with a so-called suspension-point
interferometer (SPI) that was introduced by Drever many years
ago~\cite{Drever_SPI}. While setting up an interferometer on the
suspension platform to reduce rms motion of mirrors in the main
interferometer is an idea to be used in second-generation detectors, a
true SPI is the one with an interferometer locking the upper test
masses (left panel, Figure~\ref{fig:SPI-ETMC}). Remaining noise that
appears as the differential motion of the two masses suspended from
the rigidly locked masses is mainly the common motion of the upper
stage coupled through the imbalance of the suspension systems.
An experimental demonstration has been made at a prototype
interferometer~\cite{Aso}. One option with the SPI is to use it as a
part of the xylophone. The lower-stage interferometer would be for the
low-frequency measurement with low seismic noise, and the higher-stage
interferometer would be for the high-frequency or middle-frequency
measurement with higher power.

Alternative to the long suspension will be magnetic levitation, which
is also introduced by Drever~\cite{Drever_Maglev}. Attaching magnets
to upper and lower test masses and controlling the distance of two
magnets in the vertical direction, we can trap the lower test mass,
for the horizontal directions, in a very shallow potential similar to that of  
an extremely long pendulum. The magnetic levitation might be also
a solution to reduce suspension thermal noise. It has been proposed to
use electrostatic force instead of the magnetic force so as to avoid
possible thermal noise in the magnetic system~\cite{ESD}.

Combination of the SPI and the magnetic levitation will let us tune up
the balance of the suspension systems so that the common-mode
rejection rate of the SPI can be much closer to unity. In fact, as the
purpose of the magnetic force here is to balance the suspension
systems, we can just add small magnets on test masses suspended by a
conventional pendulum.

\subsection{End-mirror cavity}\label{sec:endcavity}

One of the limiting noise sources for a second-generation detector is Brownian
motion of the coatings.
Coating Brownian thermal noise strongly depends
on the numbers of coating layers used to build the reflective
coating layer of the mirror. The higher the reflectivity of a
mirror, the higher the number of coating layers needed and the
higher the coating Brownian thermal noise.
Reduction of coating thermal noise by mechanical separation of the
first few layers and the rest was proposed by
Khalili~\cite{Khalili}. A conventional mirror is replaced by an
anti-resonant cavity that has a few coatings on the first mirror and
more coatings on the second mirror (right panel,
Figure~\ref{fig:SPI-ETMC}). The cavity is locked anti-resonant so that
temperature fluctuation in the substrate of the first mirror or
coating thermal noise of the second mirror does not matter as far as
the reflectivity of the first mirror is reasonably high.

In fact, rigid control of the end-mirror cavity allows us to further reduce the reflectivity of
 the first mirror~\cite{SomiyaETMC}. Total fluctuation of the
 end-mirror cavity is measured by an auxiliary beam and is subtracted
 from the output of the main beam in a proper way. Shot noise of the
 auxiliary beam will be imposed instead, so the power of the auxiliary
 beam should be high. While radiation pressure noise in the end-mirror
 cavity is not a problem, radiation pressure noise of the auxiliary
 beam in the main cavity cannot be suppressed by the control. As the
 number of coating layers could be zero, coating thermal noise would
 be totally replaced by quantum noise of the auxiliary beam. We could
 use one of those quantum non-demolition techniques to overcome the
 quantum limit and increase the power infinitely so that finally
 coating thermal noise could be removed without excess control noise.
It should be noted, however, that a rigid control of such a short, essentially
marginally stable cavity is known to be very difficult and in all
likelihood the noise of the alignment control system will be a
limit to the performance of such a system.

%%%%%%%%%%%%%%%%%%%%%%%%%%%%%%%%%%%%%%%%%%%%
%%%%%%%%%%%%%%%%%%%%%%%%%%%%%%%%%%%%%%%%%%%%

\section{Summary and Outlook} \label{sec:summary}
Third-generation gravitational wave detectors are envisaged to be observatories providing
a continuous stream of astrophysical meaningful data. In order to achieve this goal these detectors
need to be designed and planned to be robust machines with a sensitivity ten times better than
the advanced detectors currently under construction. In this article we have discussed how these
goals affect the design of the core instrument, in other words the optical layout of the laser interferometer(s)
of the detector. From the nature of the gravitational radiation follows that long-baseline,
L-shaped interferometers represent the ideal instruments. In order to maximise the data quality
of a single detector or of a detector integrated in a network, we found that the use of co-located
interferometers are essential. A triangular geometry of three co-located interferometers is
currently under investigation within the Einstein Telescope design study. Further constraints of the
interferometer topology and configuration are not obvious from general principles but are
expected to follow from the technical detailed of optical schemes for reducing the quantum noise and
the thermal noise in the detectors.

We have reviewed the preliminary design work on the Einstein Telescope and showed that the envisaged
sensitivity can possibly be reached
by an moderately optimistic  scaling of current technologies. In particular, we have reviewed the
noise contributions in a potential ET detector and have shown that a design featuring several
small-band interferometers might be superior over a single broadband design.
In addition, several new, advanced techniques can be used to provide a margin for further sensitivity
improvements. We have presented several such techniques, concentrating mainly  on those whose
technical readiness is sufficient for a potential integration in future large scale projects over the next
decade.

Several research programs within the world-wide gravitational wave collaborations are dedicated
to the experimental realisation and testing of  advanced techniques for third-generation detectors.
We thus expect that we will be able to significantly beat the sensitivity of the advanced detectors
which are currently under construction. This is an exciting outlook for the time when said advanced detectors
have made the first detections of gravitational waves and increasing the signal to noise ratio would enable
us to tap the full potential of the new field of gravitational wave astronomy.

\section{Acknowledgements}
This work has been supported by the Science and Technology
Facilities Council (STFC), the European Gravitational Observatory (EGO),
 the United States National
Science Foundation (NSF) and the Seventh Framework Programme (Grant
Agreement 211743) of the European Commission. K.~Somiya is supported
by Japan Society for the Promotion of Science (JSPS).



\end{document}